\documentclass{ieeeaccess}

\usepackage[utf8]{inputenc}
\usepackage{biblatex}
\usepackage{qcircuit}
\usepackage{braket}
\usepackage{amsmath,amssymb,amsmath} 
\usepackage{multirow} 
\usepackage{booktabs} 
\usepackage{graphicx} 
\usepackage{enumitem} 
\usepackage{caption} 
\usepackage{subcaption} 
\usepackage{wasysym}
\usepackage[linesnumbered,ruled,vlined]{algorithm2e}
\usepackage{listings} 
\usepackage{caption} 
\usepackage{subcaption} 
\usepackage{color}
\usepackage{hyperref}
\usepackage{textcomp}
\usepackage{algorithmic}
\newcommand{\controlxo}{ *+<.01em>{\LEFTcircle}}
\newcommand{\ctrlxo}[1]{\controlxo \qwx[#1] \qw} 
\newcommand{\qwdash}[1][-1]{\ar @{.} [0,#1]}

\newcommand{\xymultigate}[4]{*+<#1,#2>{\hphantom{#4}} \POS [0,0]="i",[0,0].[#3,0]="e",!C *{#4},"e"+UR;"e"+UL **\dir{-};"e"+DL **\dir{-};"e"+DR **\dir{-};"e"+UR **\dir{-},"i" \qw}
\newcommand{\xyghost}[3]{*+<#1,#2>{\hphantom{#3}} \qw}
\newcommand{\shortmin}{\scalebox{0.75}[1.0]{\( - \)}}

\newcommand\mylim[2]{%
    \begin{array}[t]{@{}l@{}}
     #1 \\[-1ex] \scriptstyle #2 
    \end{array}
}

\DeclareMathSymbol{\shortminus}{\mathbin}{AMSa}{"39}

\usepackage{newfloat}

\DeclareFloatingEnvironment[
  fileext = los ,
  listname = {List of circuits} ,
  name = Circuit
]{circuit}
\usepackage[capitalise]{cleveref}
\crefname{circuit}{Circ.}{Circs.} 
\Crefname{circuit}{Circuit}{Circuits} 

\newif\ifdraft
\ifdraft
\definecolor{teal}{rgb}{0,0.5,0.5}
\definecolor{violet}{rgb}{0.5,0,0.5}
\newcommand{\alnote}[1]{ {\textcolor{red} { ***Andre: #1 }}}
\newcommand{\mznote}[1]{ {\textcolor{purple} { ***Marcin: #1 }}}
\newcommand{\annenote}[1]{ {\textcolor{violet} { ***Anneriet: #1 }}}
\newcommand{\marvnote}[1]{ {\textcolor{blue} { ***Marvin: #1 }}}
\newcommand{\ewannote}[1]{ {\textcolor{cyan} { ***Ewan: #1 }}}
\newcommand{\zaidnote}[1]{ {\textcolor{teal} { ***Zaid: #1 }}}

\newcommand{\rajnote}[1]{ {\textcolor{blue} { ***Rajesh: #1 }}}
\else
\newcommand{\note}[1]{}
\newcommand{\alnote}[1]{}
\newcommand{\mznote}[1]{}
\newcommand{\annenote}[1]{}
\newcommand{\marvnote}[1]{}
\newcommand{\ewannote}[1]{}
\newcommand{\zaidnote}[1]{}
\newcommand{\rajnote}[1]{}
\fi

\def\theyear{2023}
\def\thevol{1}

\begin{document}

\history{Date of publication tbd., date of current version: Oct., 2023.}
\doi{doi}

\title{QISS: Quantum Industrial Shift Scheduling Algorithm}

\author{Anna M. Krol\href{https://orcid.org/0000-0003-0066-4299}{\includegraphics[scale=0.0035]{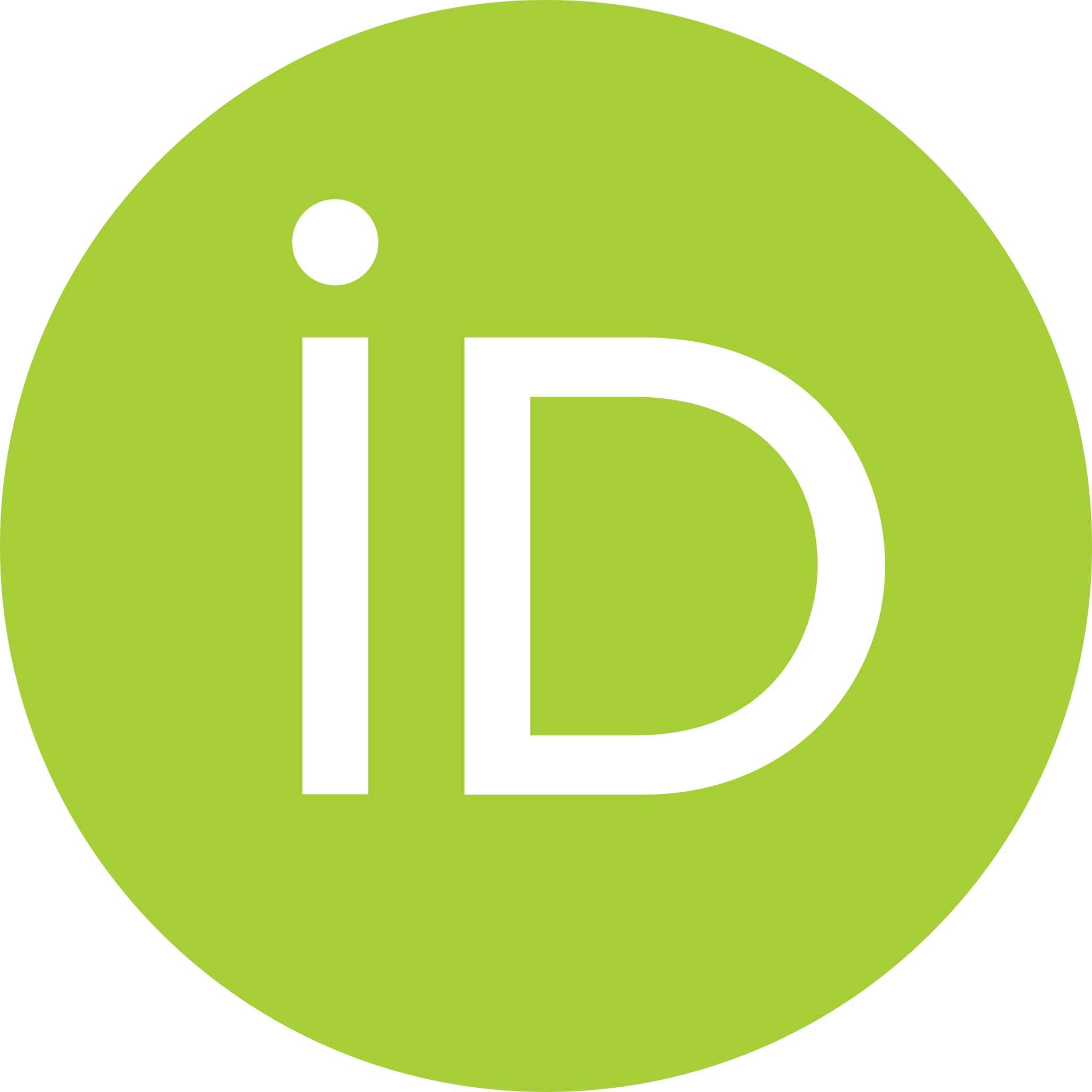}}\authorrefmark{1}, \and Marvin Erdmann\authorrefmark{2}, \and Rajesh Mishra\authorrefmark{3,$\dagger$}, \and Phattharaporn Singkanipa\authorrefmark{3,*}, \and Ewan Munro\authorrefmark{3}, \and Marcin Ziolkowski\authorrefmark{2}, \and Andre Luckow\authorrefmark{2,}\authorrefmark{4}, \and Zaid Al-Ars\authorrefmark{1}}
\date{\today}
\address[1]{Delft University of Technology, Delft, The Netherlands}
\address[2]{BMW Group, Munich, Germany}
\address[3]{Entropica Labs, Singapore}
\address[4]{Ludwig Maximilian University, Munich, Germany}
\address[$\dagger$]{Current affiliation: Department of Physics, University of Illinois Urbana-Champaign}
\address[*]{Current affiliation: Department of Physics, University of Southern California, Los Angeles, CA 90089, USA}
\corresp{Corresponding author: Anna M. Krol (email:a.m.krol@tudelft.nl)}


\begin{abstract}
In this paper, we show the design and implementation of a quantum algorithm for industrial shift scheduling (QISS), which uses Grover's adaptive search to tackle a common and important class of valuable, real-world combinatorial optimization problems. We give an explicit circuit construction of the Grover's oracle, incorporating the multiple constraints present in the problem, and detail the corresponding logical-level resource requirements. Further, we simulate the application of QISS to specific small-scale problem instances to corroborate the performance of the algorithm, and we provide an open-source repository with our code, available on \href{https://github.com/anneriet/QISS}{GitHub}. Our work shows how complex real-world industrial optimization problems can be formulated in the context of Grover’s algorithm, and paves the way towards important tasks such as physical-level resource estimation for this category of use cases.

\end{abstract}
\titlepgskip=-15pt

\maketitle


\section{Introduction} 

Many industries face complex optimization challenges too large and complex 
to solve optimally~\cite{Bayerstadler2021}. For example, in the automotive industry, these problems may include optimizing supply chain logistics~\cite{GARCIA2015153,THUN2011242,awasthi2023quantum}, quality control~\cite{FARAHANI2021106924,Hafizi_2019}, robots path planning~\cite{PhysRevApplied.18.054045}, and shift scheduling~\cite{VANDENBERGH2013367,Xie2019}. These challenges feature numerous constraints and expansive solution domains that expand exponentially with each added variable to the problem's framework.

In particular, the shift scheduling problem in automotive production networks concerns the creation of a schedule for the numerous production steps. The goal is to maximize productivity while considering dependencies between production steps, the discrete number of possible shift durations, and working regulations and preferences. As production steps, workers, and shift configurations increase, the solution space grows exponentially. 

Solving such industry-scale scenarios optimally for an extended period (e.g., one year) is not feasible, so such problems are often solved heuristically~\cite{9926440}. 
Heuristic methods can provide valid shift schedules in reasonable computation times, but cannot guarantee an optimal solution, resulting in additional costs due to over-staffing or a reduced production volume.

Quantum computing approaches for approximate solutions to such optimization problems have also been proposed~\cite{qpack_mesman}, such as~\cite{art:hybridquantumbenders,art:deepspacenetwork,art:optimizingantenna, quaser} based on quadratic unconstrained binary optimization (QUBO) and quantum annealing, as well as quantum acceleration of branch-and-bound algorithms~\cite{art:quantumbranchandbound} and reinforcement learning~\cite{art:QKSA}. 
Like classical heuristics, most of these algorithms have no mathematical bounds for the solution quality: they can provide feasible solutions under the right conditions (e.g., a structured solution space), but it is impossible to say how close to the optimal solution the algorithm got. It is therefore valuable to compare the approximate answer to that of an exact solver for problem sizes that can still be solved exactly. 


This paper provides the first implementation of an exact quantum algorithm for the industrial shift scheduling problem, which we call QISS. The term "industrial shift scheduling" is chosen for simplicity for a problem that is about the optimization of the schedule of factory shop operating hours, as described in Section~\ref{sec:sssm} and the Appendix. In other formulations of the problem one may be faced with a related version, which differs e.g. in its objective function or its constraints.

QISS is built upon the Grover's adaptive search (GAS) procedure, wherein Grover’s quantum algorithm is executed iteratively with input conditions at step $i$ derived from the output of step $i-1$. QISS then inherits the theoretical asymptotic quadratic speedup over classical algorithms for unstructured search provided by Grover’s algorithm itself.

While unstructured search is not employed in practice as a method to solve industrial-scale problems, it is often used on relatively small problem instances with the goal of benchmarking the performance of heuristic algorithms. In this context, a quadratic speedup could make a significant difference to the scope for analyzing the performance of heuristics. Whether such a form of advantage can be achieved with an exact quantum algorithm such as Grover’s may depend strongly on the use case at hand. To begin to answer the question, one must construct an explicit quantum circuit for the Grover's `oracle', which is responsible for identifying solutions that satisfy some desired search criteria.

Our work details the construction of a Grover's oracle circuit for the industrial shift scheduling problem. In particular, we show how the problem’s characteristic and complex constraints, related to production targets and intermediate storage limitations, can be incorporated directly into the oracle. We expect our work to be valuable to researchers investigating quantum computing approaches to similarly highly constrained optimization problems, and in paving the way for a thorough physical-level resource estimation for this specific problem type.

The contributions of this paper are as follows: 
\begin{itemize}
\item We formulate 
a shift scheduling problem with real-world constraints comparable to industry-relevant use cases, 
\item We implement QISS, the first 
quantum algorithm that gives an exact solution to the shift scheduling problem, and
\item We verify and evaluate 
QISS to show the correctness of the algorithm and that it can be used to solve the shift scheduling problem with quadratic speedup.
\end{itemize}

This paper starts with a background discussion in \cref{sec:background} on the industrial shift scheduling problem and current approaches to the use case, including quantum computing algorithms like Grover's search.  
The construction, implementation and validation of the quantum industrial shift scheduling algorithm are described in \cref{sec:QISS}, 
The paper concludes with \cref{sec:conclusion} with a summary of the findings. The accompanying code is publicly available on \href{https://github.com/anneriet/QISS}{GitHub}.

\section{Background} \label{sec:background}
This section will give background on the industrial shift scheduling problem, and an overview of classical and quantum solutions for the use case.

\subsection{Industrial Setting and Utility of the Problem}

The literature on shift scheduling use cases is vast due to its relevance in almost all economic and social sectors. Examples can be found for the optimization of nurse shift schedules in hospitals~\cite{Ikeda_2019}, staffing of employees in call centers~\cite{MATTIA201725}, retail~\cite{ALVAREZ2020106884,KABAK200876}, or the postal service~\cite{BARD2003745}. Overviews of more use cases and solution approaches show the whole spectrum of this optimization problem~\cite{syslitrev,VANDENBERGH2013367}.

Shift scheduling is also essential to production planning in manufacturing facilities, especially in the automotive industry. 
To automate the scheduling process and minimize costs while maximizing productivity, manufacturers use optimization algorithms for shift scheduling. An optimization algorithm can analyze vast amounts of data and variables, such as the production schedule, worker availability, working regulations, and dependencies, to generate a shift schedule that meets production targets, reduces labor costs, and improves worker satisfaction.

The main difference between shift scheduling in the manufacturing sector and other sectors is the added constraint of a target production volume. 
Unlike other versions of the shift scheduling problem, the objective is not to cover all shifts adequately. Instead, the goal is to maximize the overall productivity of end-to-end manufacturing systems comprising multiple production sites, referred to as \textit{shops}, while minimizing labor costs and complying with legal regulations.

One of the critical business motivations for implementing an automated shift scheduling algorithm is the significant problem size associated with industrial manufacturing facilities. Companies operate in global production networks, and each facility has its own unique production lines, teams, and shifts, making the scheduling process complex and challenging. The optimal scheduling of thousands of employees working in various roles, including production, maintenance, logistics, and administration, is highly complex and implies a high potential for reduced labor costs and more consistent production volumes.


More reliable production volumes are a crucial performance metric 
for manufacturing companies, especially in the automotive industry, and any disruptions to the production process can negatively affect this metric. An optimal schedule ensures that shifts are planned prudently so that an adequate number of workers is always available. Interdependent steps in the production process must be synchronized, and the limited intermediate storage between these steps, so-called \textit{buffers}, must be managed carefully to minimize production downtime and increase the probability of reaching the production target corridor.

Depending on the industry sector and the manufacturing entity's size, the problem's scale can vary tremendously. In the automotive industry, the annual production volumes of individual vehicle models reach values of several hundred thousand units with small margins of a few percent of the overall volume. 



The shift scheduling problem is known to be NP-complete \cite{Karp1972}. The number of different shift configurations scales exponentially with the number of production steps in the process and the number of days in the schedule. 
Thus, the solution space of industry-scale scenarios with around ten production steps and approximately 300 working days per year is too large to be fully explored with classical methods
in a reasonable amount of time, even when considering working time restrictions and regulative constraints cutting the feasible solution space considerably. 

In the remainder of this paper, unless otherwise stated, when we refer to the scheduling problem, this means the volume-constrained industrial shift scheduling problem.

\subsection{Classical Heuristics}
Because of its complexity and large scale in industrial applications, the volume-constrained shift scheduling optimization problem can currently only be solved heuristically. Constructive heuristics allow dividing the problem into smaller parts, solving each of them individually, updating the constraints according to the solutions found for the prior parts, and combining them in the end to a valid solution~\cite{rocha2014constructive}.

The combined solutions of the subproblems do not necessarily form a globally optimal solution for the entire problem, even if each individual subproblem can be solved optimally. An example of such a constructive heuristic based on optimal solutions of subproblems is the following: a branch-and-cut algorithm identifies the optimal shift schedule for the first week of one year~\cite{rocha2014constructive}. The solution for this subproblem is used as the basis for the subsequent week, the constraints and restrictions are adjusted accordingly, and the branch-and-cut algorithm searches for the optimal solution for the next week.  

This process is iteratively repeated until the algorithm is unable to identify a valid solution for a given week. In this case, the search procedure is repeated for the last week with a valid solution. The previously found optimal solution is penalized such that another valid solution is used as a new basis for the subsequent week. This constructive search procedure runs until 365 consecutive days with valid schedules are found.

Even such a heuristic search procedure can take several hours of computation time, due to the unstructured nature of the solution space of the shift scheduling problem. Unstructured solution spaces are difficult to explore for optimization algorithms, because solutions with small differences in their configuration can produce vastly different objective function values. Therefore, procedures like branch-and-cut algorithms - which are based on excluding portions of the solution space that are guaranteed to not include a solution that is better than the worst possible solution in the rest of the solution space - scale badly with the size of shift scheduling problems.

The optimization of shift schedules has long been known to be hard to solve optimally. Therefore, many other heuristic algorithms have been explored for different versions of the use case in literature, such as simulated annealing, genetic algorithms~\cite{bailey1997using}, tabu search~\cite{DOWSLAND1998393}, and ant-colony optimization~\cite{GUTJAHR2007642}. The employee shift scheduling problem has also been identified as an interesting use case for evaluating quantum computing algorithms and comparing their performances to classical methods~\cite{luckow2021quantum,bayerstadler2021industry}. However, none of these studies has considered the volume-constrained shift scheduling use case, which is significantly more complex to model.

\subsection{Grover's Algorithm} \label{sec:groversbg}
QISS uses Grover's adaptive search~\cite{art:PASwithgrovers,art:GASforCPBO} to find optimal shift schedules, which combines classical adaptive search with Grover's search algorithm. A short introduction to Grover's search is given here, while the full algorithm is described in \cref{sec:QISS}.

Grover's algorithm 
was initially formulated as a search algorithm for unsorted databases, but has since then been used in various applications, ranging from optimization~\cite{art:GASforCPBO} to approximate search~\cite{qibam}. The algorithm is designed for cases where only one record satisfies a particular property. Any classical algorithm must take $O(N)$ steps because, on average, half of the $N$ records need to be evaluated to find the target record. A quantum computer can identify the record in only $O(\sqrt{N})$ steps with Grover's algorithm, thereby giving an asymptotic quadratic speedup over a classical random search algorithm~\cite{art:groversalg}.

A high-level overview of the circuit implementation of Grover's algorithm is shown in \cref{fig:groversoracle}. Grover's algorithm consists of the following steps~\cite{art:GASforCPBO,art:groversalg}:
\begin{enumerate}
    \item Initialization of the system to an equal superposition of all states in the solution space. This is done by applying a Hadamard gate to each qubit. 
    \item An oracle that recognizes the states that are valid solutions, and multiplies their amplitudes by -1. The oracle of QISS is implemented as $O^\dagger$CNOT$(\ket{-})O$, where the CNOT gate is controlled by condition qubits in $\ket{q_c}$, which are set to $\ket{1}$ if all conditions are fulfilled. 
    \item Grover's diffusion operator $D$, which multiplies the amplitude of the $\ket{0}_n$ state (or all states except $\ket{0}_n$) by -1. This has the effect of inverting all amplitudes in the quantum state about the mean, which amplifies the magnitudes of all states of interest and decreases the magnitudes of all other states. 
    \item These are combined into Grover's rotation operator {$G = D O^\dagger$CNOT$(\ket{-})O$}. 
    \item The rotation operator can be repeated to amplify the state(s) of interest. A Grover's search with $j$ rotations has $j$ repeated applications of rotation operator $G$. 
    \item Sample the resulting state. For a problem with a solution space of size $N$ and $t$ valid solutions, the probability of measuring a valid solution after $j$ rotations is $P=t\cdot|k_j|^2$, where $k_j = \frac{1}{\sqrt{t}}\sin\left((2j+1)\cdot\theta\right)$ and $\sin^2\theta = \frac{t}{N}$.
\end{enumerate}

Maximum amplification of the states of interest occurs at $j = \frac{\pi}{4}\sqrt{\frac{N}{t}}$, where $N$ is the size of the solution space, and $t$ is the number of valid solutions.


\begin{circuit}[htbp]
\[
\footnotesize
\begin{array}{c}
\Qcircuit @C=0.5em @R=.5em @!R{
\lstick{\ket{q_s}=\ket{0}} & / \qw & \gate{H} & \qw &  \multigate{3}{Oracle} & \qw & \multigate{3}{\parbox{3.5em}{\centering $Reset=$ \\ $Oracle^\dagger$}} & \gate{\mathit{Diffuser}} & \qw & \meter &  \\
\lstick{\ket{q_b}=\ket{0}} & / \qw & \qw & \qw & \ghost{Oracle} & \qw & \ghost{\parbox{3.5em}{$Oracle^\dagger$}} & \qw & \qw & \qw \cwx  & \qw \\
\lstick{\ket{q_\$}=\ket{0}} & / \qw & \qw & \qw & \ghost{Oracle} & \qw & \ghost{\parbox{3.5em}{$Oracle^\dagger$}} & \qw & \qw & \qw \cwx  & \qw \\
\lstick{\ket{c}=\ket{0}} & / \qw & \qw & \qw & \ghost{Oracle} & \ctrl{1} & \ghost{\parbox{3.5em}{$Oracle^\dagger$}} & \qw & \qw & \qw \cwx  & \qw \\
\lstick{\ket{-}} & \qw & \qw & \qw & \qw & \targ & \qw & \qw  & \qw & \qw \cwx  & \qw  \gategroup{1}{5}{5}{8}{1.5em}{--} \\
\lstick{\textrm{Class. bits}}& \cw & \cw & \cw & \cw & \cw & \cw & \cw & \cw & \cw \cwx  &  \rstick{\textrm{Output}}\cw
}
\end{array}\]
\caption{Circuit for Grover's algorithm as used in QISS. Within the dashed line is Grover's operator $G$, which can be repeated to maximize the amplification of the target states. \rajnote{After reading the rest of the paper, I feel that having a generic Grover's circuit here is better.$\ket{q_s}$, $\ket{q_b}$ and $\ket{q_\$}$ have not been defined yet. The reader might get confused. Also, the condition qubits are referred to as $\ket{q_c}$ instead of $\ket{c}$ in step 2.}}
    \label{fig:groversoracle}
\end{circuit}

\section{Algorithm design and validation} \label{sec:QISS}

QISS leverages Grover's algorithm~\cite{art:tightboundsonquantumsearching} in combination with an adaptive search procedure, collectively referred to as Grover's adaptive search (GAS)~\cite{art:PASwithgrovers,art:GASforCPBO}, to provide exact solutions to the industrial shift scheduling problem. GAS is described in \cref{alg:GAS}.

QISS returns shift schedules that satisfy the problem constraints, with the lowest cost. Compared to a classical brute-force search, QISS benefits from the asymptotic quadratic speedup delivered by Grover's algorithm. As we remarked in the introduction, this speedup may be useful in extending the scope for benchmarking the performance of heuristic algorithms designed to tackle industrial-size problem instances.

We use a simplified model of a vehicle assembly line for this proof-of-concept of QISS (\cref{sec:sssm}). Based on the constraints of the model, we construct an oracle that marks input states (schedules) as valid or invalid. 
First, we map all possible combinations of possible shift lengths to binary numbers in equal superposition on qubit register $\ket{q_s}$. This superposition of states defines the solution space, and the qubits in this register are the only qubits that are measured at the end (\cref{subsec:dataencoding}). 

Then we define qubit registers to calculate the buffer content, total cost and solution validity (\cref{subsec:qubitregs}). 
Calculating buffer content and cost is done through adders in the Fourier basis, using the technique from~\cite{art:additiononqc}.\alnote{can give a brief intuition of how this works? What does it mean in terms of \# operations?} Quantum Fourier Transforms are used to transform the qubits from the Fourier basis to the computational basis and back (\cref{subsec:adding}). 

Conditions are checked by addition or subtraction of relevant upper or lower limit values, so a condition is (not) met if the qubits hold a (positive) negative number. This can be easily checked because we use Two's complement encoding (\cref{subsec:constraintchecks}).

To mark the states we want to amplify, we use a register $\ket{c}$ with a condition qubit for each of the conditions of the model. 
The corresponding qubit is set to $\ket{1}$ when a condition is met. Using a multi-controlled NOT gate, a single output qubit is flipped if all condition qubits are in $\ket{1}$. When the output qubit is initialized to the $\ket{-}$ state, this operation applies a phase shift to all the states that we want to mark. We verify the completed circuit for the first day (\cref{subsec:completedcircuit}), and then extend it for scheduling multiple days (\cref{subsec:extending}).

By updating the maximum allowed cost of the solution, only those schedules with lower cost are amplified by Grover's algorithm. When there are no more valid solutions left, the optimal solution has been found(\cref{sec:costconstraint}). For bigger problem sizes, we do not know the number of valid solutions nor do we know if we have actually reached the optimal solution. This is why we need a general stop condition. 

With GAS, the maximum cost at each iteration is set to the lowest cost that has been found, up to that point. We will stop the algorithm when the total number of rotations exceeds $\sqrt{N}$. This yields goods results, which we have validated using simulations for shift schedules up to three days (\cref{GASstopcondition}).

\begin{algorithm}[htbp]
\caption{\label{alg:GAS} Grover's adaptive search} 
\DontPrintSemicolon
\small
\SetKw{KwSet}{Set}
\SetKw{KwGoal}{Goal:}
\KwGoal{} Find the lowest value for $y = f(x)$\;
\KwIn{Oracle that flags all states $x$ where $f(x) < y_{min}$}
\KwSet{} m to 1 and $\lambda$ to 6/5 or any value $1 < \lambda \leq 4/3$\;
\KwSet{} $x_{best}$ to an initial (valid) state\;
\KwSet{} $y_{min}$ to a best guess for the solution\;
\Repeat{\upshape{a stop condition is met}}{
    Choose j randomly from all integers $0 \leq j < m$\;
    Apply Grover's search with $j$ rotations and measure the circuit\;
    \KwSet{} $x$ to the measurement result\;
    \If{$f(x) < y_{min}$}{
        \KwSet{} $y_{min}$ to $f(x)$ and $x_{best}$ to $x$\;
        \KwSet{} $m$ to 1\;
        }
    \Else{
        \KwSet{} m to the smaller of $\lambda \cdot m$ and $\sqrt{N}$, where $N$ equals the total number of possible solutions
        }
}
\KwOut{$x_{best}$ and $y_{min}$}
\end{algorithm}

\subsection{Simplified Model for Shift Scheduling} \label{sec:sssm}

\begin{figure*}
    \centering
    \includegraphics[width=\linewidth]{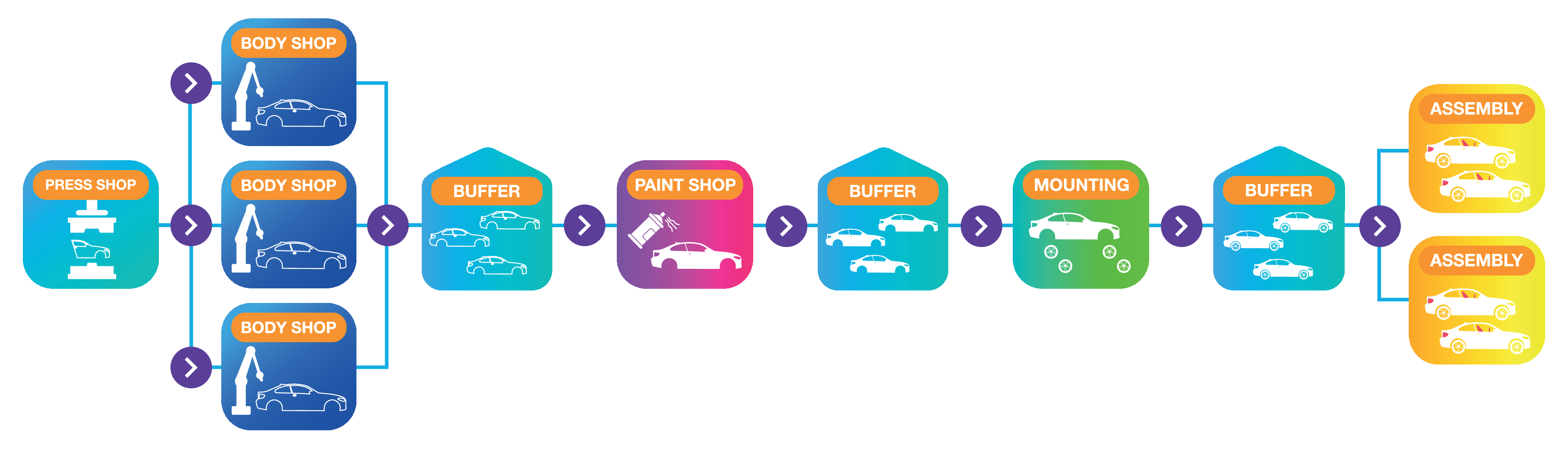}
    \caption{An example of an automotive productive line with five steps: a press shop, three parallel body shops, a paint shop, mounting and two separate assemblies. There are buffers between the body, paint, mounting,  and assembly steps.}
    \label{fig:FSM}
\end{figure*}

The production process of vehicles is a series of consecutive steps, 
some of which are parallelized to increase productivity. In \cref{fig:FSM}, a sequence of five steps is depicted: a \textit{press shop}, \textit{body shops}, a \textit{paint shop}, \textit{mounting}, and two \textit{assemblies}. Here, a singular \textit{press shop} supplies parts to three concurrent \textit{body shops} that assemble the vehicle's body. These assemblies then progress to a unified \textit{paint shop} for coloring, followed by \textit{mounting} where the vehicle's body is joined to the chassis (also referred to ``marriage''). The production culminates in the \textit{assembly}, bifurcated into two parallel processes in this example. It is noteworthy that vehicles are buffered between certain stages to modulate throughput.

\begin{figure}
    \centering
    \includegraphics[width=0.8\linewidth]{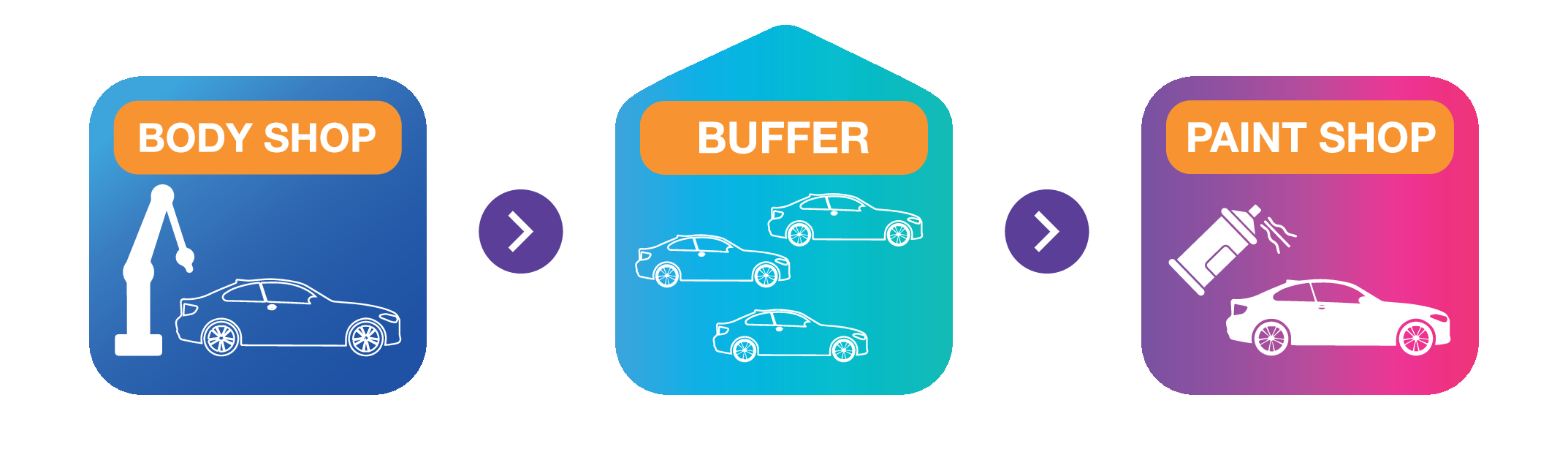}
    \caption{The structure of the simplified model: a body shop and a paint shop that share a buffer.}
    \label{fig:SSSM}
\end{figure}

This paper will consider a simplified model of the automotive production line consisting of just two shops, a body shop and a paint shop, with a single shared buffer between them. The goal is to find the optimal working time for each shop to meet an (annual) production volume target with minimal costs. 
The model is shown schematically in \cref{fig:SSSM}. 

QISS outputs a schedule with a chosen shift length for each shop for each day. A solution is \textit{valid} if the schedule does not violate any constraints.The objective is then to find the \textit{cheapest} valid solution, where the sum of the operating costs of the shops for the chosen shift lengths is as low as possible. 
Shifts are not assigned for individual employees, but at the level of shops: how many hours should the whole shop operate on a given day. 


The simplified model has the following features and constraints:
\begin{itemize}
    \item 2 shops: A body shop (S1) and a paint shop (S2).
    \item Both shops produce 1 unit/hour at \$1/hour.
    \item Maximum of 1 shift per day per shop.
    \item Allowed shift lengths (in hours) for the body shop are [0, 5, 8, 10] and for the paint shop they are [0, 4, 7, 9].
    \item No additional costs for different shift assignments on consecutive days.
    \item The shared buffer has an initial content of 5 units.
    \item The buffer cannot hold less than 0 or more than 10 units at the end of each day.
    \item Target output volume $V^* = 8n$ for $n$ days.
    \item Over- or underproduction of $\Delta = 0.05V^*$ is allowed: actual output volume is $V = 8n \pm \Delta$.
\end{itemize}

The parameters of our simplified model have been chosen to ensure the problem cannot be solved too trivially or easily. Scenarios in which the problem may become simple to solve include those where: the possible shift lengths for shops S1 and S2 are identical; the target output volume is so high (low) that the shops must essentially work the maximum (minimum) number of hours; the buffer capacity is large enough to accommodate multiple days' worth of output from shop S1. We further remark that the conditions we have chosen are realistic, e.g. it would not be practical or cost-effective to construct a buffer capable of storing multiple days' worth of output stock, which could run into thousands of vehicles for real-world scenarios.

A mathematical formulation of our objective function to minimize is given by
\begin{align}
    C_{f} = \sum_{i=1}^{\mathit{\#shifts}} \sum_{j=1}^{\mathit{\#shops}} S_{ij}C_j,
\end{align}
where \textit{\#shifts} is the product of the total number of days and the number of shifts per day, \textit{\#shops} is the total number of shops, $S_{ij}$ denotes the assigned shift hours for shift $i$ for shop $j$  and $C_j$ denotes the cost of operation per hour for shop $j$. The buffer is assumed to have a constant cost and does not contribute to the cost function. 

For this simplified model, the number of possible schedules grows with $16^n$ for $n$ days, and a year with 280 working days has $2^{1120}$ possible schedules. Checking all of these schedules for validity and finding the cheapest of all the valid solutions is a computationally prohibitive task. In the Appendix we provide supplementary information on the simplified model, and discuss some of the complexities that arise when considering more sophisticated, real-world models.
  

\subsection{Data encoding} \label{subsec:dataencoding}
The shift assignments per shop function as the decision variables and are encoded in binary form, using $\log_2(i)$ qubits to store $i$ different choices for shift assignments. Compared to a one-hot encoding, this approach uses fewer qubits and ensures that each shop is assigned a single shift length per day. 
The encoding for the shift length 
for each shop in the simplified model is shown in \cref{tab:shop1shop2encoding}. We need four qubits per day, two for each shop. Shift lengths are combined as $\ket{S1}\ket{S2}$, for example $\ket{01}\ket{11}$ means a shift of 5 hours for shop 1 and a shift of 9 hours for shop 2. Encoding for multiple days is done by simply adding four qubits per day to allow for solutions with different shift lengths for different days. 

\begin{table}[htbp]
\caption{Encoding of shift length in hours into qubit states for shops 1 and 2.
} \label{tab:shop1shop2encoding}
\centering
\begin{tabular}{ccc}
\toprule
 & \multicolumn{2}{l}{Shift length in hours} \\  \cline{2-3} \addlinespace[0.2em] 
 \multirow{-2}{*}{Qubit state}                             & Shop 1   & Shop 2    \\  \midrule
$\ket{00}$                   & 0        & 0         \\ 
$\ket{01}$                   & 5         & 4         \\ 
$\ket{10}$                   & 8         & 7         \\ 
$\ket{11}$                   & 10        & 9         \\ \bottomrule
\end{tabular}
\end{table}

The combination of these states for both shops and all days will make up the solution space for QISS. 

The number of units in the buffer is stored in a separate register in Two's complement form. Compared to other signed number representations, Two's complement has the advantage that the operations for addition, subtraction, and multiplication are identical to those for unsigned binary numbers. With Two's complement, the most significant bit (MSB) of a binary number is `0' for positive numbers and zero, and `1' for negative numbers, which simplifies checking for the constraints later. 
To convert a negative number to its $b$-bit Two's complement equivalent, we can add $2^b$ and convert the resulting number as if it were an unsigned binary number. For example, the number $-2$ in three-bit Two's complement is $2^3-2=8-2=6=$`110'. 

\begin{circuit*}[hbt]

\[
\begin{array}{c}
\small
\Qcircuit @C=1em @R=.0em  @!R{ 
\lstick{\ket{q_{S1,1}}} & \qw & \ctrl{2} & \ctrl{3} & \ctrl{4}  & \ctrl{5} & \ctrl{6} & \qw & & & \qw & \ctrl{2} & \qw \\
\lstick{\ket{q_{S1,2}}} & \gate{X} & \ctrl{1} & \ctrl{2} & \ctrl{3}  & \ctrl{4} & \ctrl{5} & \qw &\push{\rule{0em}{1.5em}} & &\gate{X} & \ctrl{1} & \qw \\
\lstick{\ket{q_{b0}}} & \qw & \gate{P(\frac{2\pi}{2}8)} & \qw & \qw & \qw & \qw & \qw & &  &\qw & \multigate{4}{U(8)} & \qw \\
\lstick{\ket{q_{b1}}} & \qw & \qw & \gate{P(\frac{2\pi}{4}8)} & \qw & \qw & \qw & \qw & \push{\rule{0em}{0em}=\rule{0em}{0em}} & & \qw & \ghost{U(8)} & \qw \\
\lstick{\ket{q_{b2}}} & \qw & \qw & \qw & \gate{P(\frac{2\pi}{8}8)} & \qw & \qw & \qw &  & & \qw & \ghost{U(8)} & \qw \\
\lstick{\ket{q_{b3}}} & \qw & \qw & \qw & \qw & \gate{P(\frac{2\pi}{16}8)} & \qw & \qw & & & \qw & \ghost{U(8)} & \qw \\
\lstick{\ket{q_{b4}}} & \qw & \qw & \qw & \qw & \qw & \gate{P(\frac{2\pi}{32}8)} & \qw & & & \qw & \ghost{U(8)} & \qw 
}
\end{array}
\]
\caption{Conditional data encoding of the number 8 on qubit vector $\ket{q_b} = \ket{q_{b0}q_{b1}q_{b2}q_{b3}q_{b4}}$. In future circuits, we will refer to this type of circuit as the controlled gate $U(a)$ on the right for the encoding of the number $a$.}
    \label{fig:from21}
\end{circuit*}

We use circuits like the one in \cref{fig:from21}~\cite{art:additiononqc} to encode the number of units produced per shift. This circuit encodes the decimal value 8 into the buffer qubit register $\ket{q_b}$ when the decision variable register $\ket{q_s}$ is in state $\ket{10}$. This corresponds to the 8 units produced during the 8-hour shift of Shop 1. 

The phase gates $P(\lambda)$ in this circuit apply a rotation around the Z-axis by angle $\lambda$ to the qubit. If the phase gate is controlled by one or multiple other qubits, it only applies the rotation when all controlling qubits are $\ket{1}$. 
The matrix representation for this gate is:
\begin{align}
    P(\lambda) = \begin{bmatrix} 1 & 0 \\ 0 & e^{i\lambda} \end{bmatrix} \label{eq:plambda}
\end{align}

This can be used to encode the number of units produced for each possible shift length, which is shown in \cref{fig:24}.

\begin{circuit}[hbt]
\[
\begin{array}{c}
\small
\Qcircuit @C=0.3em @R=.5em {
 & \ctrl{1} & \qw & \ctrl{1} & \gate{X} & \ctrl{1} & \qw & \ctrl{1} & \qw  & & & \ctrlxo{1} & \qw \\
 & \ctrl{1} & \gate{X} & \ctrl{1} & \qw & \ctrl{1} & \gate{X} & \ctrl{1} & \qw & \push{\rule{0em}{0em}=\rule{0em}{0em}} & & \ctrlxo{1} & \qw\\
\lstick{\ket{q_b}} & \gate{U(10)} & \qw  & \gate{U(8)} & \qw  & \gate{U(0)} & \qw  & \gate{U(5)} & \qw &&&\gate{S1}& \qw \inputgroupv{1}{2}{1em}{1em}{\ket{q_{S1}}} 
}    
\end{array}
\]
\caption{Encoding the number of units produced for the body shop (S1) in superposition onto $\ket{q_b}$, using the U(a) gate from \cref{fig:from21}.}
    \label{fig:24}
\end{circuit}

\subsection{Qubit registers} \label{subsec:qubitregs}

\begin{table}
\caption{Description of the qubit registers used in QISS, including the number of qubits in each.} \label{tab:qubitregisters}
\setlength{\tabcolsep}{4pt}
\begin{tabular}{rllcl}
\toprule
\multicolumn{2}{l}{Register}                  & Description         &Qubits: 1 day & $n$ days   \\ \midrule
\multirow{2}{*}{$\ket{q_s}$} & $\ket{q_{S1}}$ & Shop 1              & 2& $2n$                            \\
                             & $\ket{q_{S2}}$ & Shop 2              & 2 & $2n$                            \\
\multicolumn{2}{r}{$\ket{q_b}$}               & Buffer   & 5 & 6 \\
\multicolumn{2}{r}{$\ket{a}$}           & Ancilla & 5& $6n-1$                            \\
\multicolumn{2}{r}{$\ket{q_\$}$ }                & Cost                 & 5& $\left\lceil{\log_2(19n)}\right\rceil$      \\
\multicolumn{2}{r}{$\ket{c}$}               & Conditions          & 4 & $n+3$                           \\
\multicolumn{2}{r}{$\ket{-}$}                 & Mark states & 1& 1          \\ \midrule
 & & Total & 24& $11n+9+\left\lceil{\log_2(19n)}\right\rceil$ \\ \bottomrule                   
\end{tabular}
\end{table}
QISS uses six distinct qubit registers, which are the following:
\begin{enumerate}
    \item \textbf{Decision variables} $\ket{q_{S1}}$ and $\ket{q_{S2}}$: This register stores the assigned shifts to each shop. For $i$ different possible shift assignments per shop, we need $\lceil \log_2(i)\rceil$ qubits. Each shop has four possible choices of shift assignment, so we need $\left\lceil{\log_2(4)}\right \rceil = 2$ qubits per shop per day. For $n$ days, we will need  $\lceil n \cdot \log_2(i)\rceil$ qubits.
    \item \textbf{Buffer occupancy} $\ket{q_b}$: This register stores the number of units in the buffer after a given day's work. It must contain enough qubits to capture buffer overflows, which correspond to infeasible solutions. For the simplified model, this is the case when the buffer is already filled to the maximum allowed value $B_{max}$, and the next day, shop 1 produces the maximum number of units while shop 2 does not consume any units. This leads to a buffer occupancy of $10+10=20$ units, which requires $\left\lceil{\log_2(10+10)}\right\rceil +1 = 6$ qubits. This is the maximum buffer occupancy that we need to calculate accurately, since there is no reason to have accurate computations of buffer occupancy for subsequent days after the solution has already been marked infeasible. 
    \item \textbf{Ancilla qubits} $\ket{a}$:  This register is used to impose a floor of zero on the buffer occupancy, since it cannot contain a negative number of vehicles. For every day that needs to be scheduled, the $MAX(0,\tilde{B})$ operation described in \cref{subsec:adding} requires as many ancilla qubits as there are qubits in the buffer register, which is $6n$ for $n$ days for the simplified model. On the first day, only 5 qubits are required, because the buffer value cannot exceed $5+10 = 15$ units. This can be taken advantage of by using a modified $MAX(0,\tilde{B})$ circuit. 
    \item \textbf{Cost qubits} $\ket{q_\$}$: This register records the operational cost of implementing a shift schedule. It must contain enough qubits to store the maximum possible cost without overflowing. This can be done as an unsigned number because no negative costs are possible. The maximum cost per day for the simplified model is \$19, so for $n$ days we need a cost register of  $\left\lceil{\log_2(19n)}\right\rceil$ qubits.
    \item \textbf{Condition qubits} $\ket{c}$:  
    This register keeps track of all the different constraints, and whether they are satisfied or violated. For each constraint, we thus require one qubit to store a boolean value. Three qubits are needed for one day, and we need $n+2$ qubits for $n$ days.
    \item \textbf{Marking of valid states} $\ket{-}$:  
    A single qubit that is initialized in the state $\ket{-}$. By applying a multi-controlled not gate from the condition qubits to this output qubit, we apply a phase of -1 to all valid solutions, as indicated by all the constraints being met. The subsequent operations of Grover's algorithm will amplify the amplitude of these marked valid solutions.
\end{enumerate}
 An overview of the number of qubits used by the oracle can also be found in 
\cref{tab:qubitregisters}. 
In total, QISS uses 18 qubits to find the optimal schedule for a single day. For $n$ days the required number of qubits is $11n+8$, or $11n+9$ when there is no separate implementation of a modified $MAX(0,\tilde{B})$ for the first day.

\subsection{Calculating buffer values} \label{subsec:adding}
We will need to perform addition and subtraction to calculate the value of the buffer occupancy on each day, and the total output volume produced by the shops. We can apply the relevant $SX$ or $U(X)$ gates to the target qubits, as shown in \cref{fig:alternative22}. 

\begin{circuit}[htbp]
\[
\begin{array}{c}
\small
\Qcircuit @C=0.5em @R=0.5em  {
 \lstick{\ket{q_{S1}}} & \ctrlxo{2} & \qw & \qw & \qw\\
 \lstick{\ket{q_{S2}}} & \qw & \ctrlxo{1} & \qw & \qw \\
\lstick{\ket{q_b}} & \gate{S1} & \gate{S2}& \gate{U(\shortminus C)} & \qw 
} 
\end{array}
\]
\caption{Calculating $S1 + S2 - C$}
    \label{fig:alternative22}
\end{circuit}

Rather than defining a separate subtraction method, we negate the value to be subtracted, and subsequently use the addition method described in Section \ref{subsec:dataencoding}. The negative of the value $A$ is calculated as $(-1)*A = 2^m-A$ for a register with $m$ (qu)bits using Two's complement. For values already encoded in the quantum circuit, 
the negation of the bits is done by applying an X-gate to all the qubits in the computation basis, and then adding 1. 
This circuit is shown in \cref{fig:negationcircuit}.

\begin{circuit}[htbp]
\[
\begin{array}{c}
\small
\Qcircuit @C=0.5em @R=0.5em  {
 \lstick{\ket{q_{S1}}} & \ctrlxo{1} & \qw & \qw & \qw & \qw & \qw &&& \ctrlxo{1} & \qw\\
\lstick{\ket{q_b}} & \gate{S1} & \gate{QFT^\dagger} & \gate{X}& \gate{QFT} & \gate{U(1)} & \qw &\push{\rule{0em}{0em}=\rule{0em}{0em}} && \gate{\shortminus S1} & \qw
}
\end{array}
\]
\caption{Calculating $\ket{q_b} = -S1$}
    \label{fig:negationcircuit}
\end{circuit}

Before constructing the complete circuit, we need to define a final building block. This is the $MAX(0,\tilde{B})$ operation, which enforces that the buffer always holds zero or more units. If the scheduled shop hours lead to less than zero units in the buffer, the number of units in the buffer is instead set to zero.
The circuit to calculate $MAX(0,\tilde{B})$ is shown in \cref{fig:max0B}. 

After calculating the buffer occupancy, we must transform the register back to the computational basis for the application of the $MAX(0,\tilde{B})$ operation and the constraint checks. Transformations from the computational to the Fourier basis is done with Quantum Fourier Transforms (QFT), and to return to the computational basis an inverse QFT is used. 

As we are encoding the buffer occupancy value in Two's complement, we can use the MSB ($\ket{q_{b0}}$) to check if the number is smaller than zero. If that is the case, we set all the other qubits to zero. We do this for each qubit in turn, by setting an ancilla qubit to $\ket{1}$ if both the MSB and the current target bit are in state $\ket{1}$ with a Toffoli gate. Then, we apply a CNOT from an ancilla in the $\ket{0}$ state to the current target bit, setting it to zero. 
The inverse oracle (Oracle$^\dagger$) is also used in the circuit (see \cref{fig:groversoracle}),
so a "fresh" ancilla needs to be used for each bit of the buffer register. 
The circuit requires $b$ qubits per day for the $b$ qubits in buffer register $\ket{q_b}$.

\begin{circuit}[hbt]
\[
\begin{array}{c}
\small
\Qcircuit @C=0.3em @R=0.5em {
 \push{\ket{q_{b0}}}& & \ctrl{1} & \qw & \qw & \ctrl{2} & \qw & \qw & \ctrl{3} & \qw & \qw & \ctrl{4} & \qw & \qw & \ctrl{5} & \targ & \qw &&& \multigate{5}{\parbox{2.5em}{$MAX$\\ $(0,\tilde{B})$}} & \qw\\
 \push{\ket{q_{b1}}} && \ctrl{4} & \targ & \qw & \qw & \qw & \qw & \qw & \qw & \qw & \qw & \qw & \qw & \qw & \qw & \qw &&& \ghost{\parbox{2.5em}{$MAX$}} & \qw\\
 \push{\ket{q_{b2}}} && \qw & \qw & \qw & \ctrl{3} & \targ & \qw & \qw & \qw & \qw & \qw & \qw & \qw & \qw & \qw & \qw &&& \ghost{\parbox{2.5em}{$MAX$}} & \qw\\
 \push{\ket{q_{b3}}} && \qw & \qw & \qw & \qw & \qw & \qw & \ctrl{2} & \targ & \qw & \qw & \qw & \qw & \qw & \qw & \qw & \push{\rule{0em}{0em}=\rule{0em}{0em}} && \ghost{\parbox{2.5em}{$MAX$}} & \qw\\
 \push{\ket{q_{b4}}} && \qw & \qw & \qw & \qw & \qw & \qw  & \qw & \qw & \qw & \ctrl{1} & \targ & \qw & \qw & \qw & \qw &&& \ghost{\parbox{2.5em}{$MAX$}} & \qw\\
\push{\ket{0}}&& \targ & \ctrl{-4} \qwx & \push{\ket{0}} \qw & \targ & \ctrl{-3} \qwx & \push{\ket{0}} \qw& \targ & \ctrl{-2} \qwx & \push{\ket{0}} \qw & \targ & \ctrl{-1} \qwx & \push{\ket{0}} \qw  & \targ & \ctrl{-5}& \qw &&& \ghost{\parbox{2.5em}{$MAX$}} & \qw\\
}
\end{array}
\]
\caption{Circuit for forcing the value of $\ket{q_b}$ to be greater than or equal to zero, when $\ket{q_b}$ is stored in Two's complement. The $\ket{0}$ in the circuit means that the ancilla qubit is reset to zero, or a "fresh" ancilla in the $\ket{0}$ is used. The reset operation involves non-reversible measurement operations, but the circuit can be made fully reversible with the additional ancilla qubits. This circuit will be referred to as $MAX(0,\tilde{B})$.}
    \label{fig:max0B}
\end{circuit}

The final value for the buffer $B_{out}$ after each day can be calculated as:
\begin{align*}
    \tilde{B} &= B_{init} + S1 - S2\\
    B_{out} &= \mathit{MAX}(0,\tilde{B})
\end{align*}

This circuit is shown in \cref{fig:Boutoneday}. The value of $-S2$ can be implemented from $S2$ as in \cref{fig:negationcircuit}, however, since $S2$ is generated from the classical input to the circuit, it can also be implemented without using additional gates. This is done by replacing all the shift lengths ($s_i$) for S2 with their corresponding $2^m-s_i$ and encoding the circuit $2^m-S2$ with these new values as in \cref{fig:24}. Because we are using Two's complement, this is equivalent to using $-S2$. Directly negating the input values uses fewer gates than the method in \cref{fig:negationcircuit}, and is what we will use for the rest of the circuits.

\begin{circuit}[hbt]
\[
\begin{array}{c}
\small
\setlength\fboxsep{0pt}
\Qcircuit @C=0.5em @R=0.5em {
\lstick{\ket{q_{S1}}} & \qw & \ctrlxo{2} & \qw & \qw & \qw& \qw & \qw 
&&& \qw &\ctrlxo{1} & \qw\\
\lstick{\ket{q_{S2}}} & \qw & \qw & \ctrlxo{1} & \qw & \qw& \qw & \qw 
&&& \qw &\ctrlxo{1} & \qw\\
\lstick{\ket{q_{b}}} & \gate{\mylim{B}{init}}  & \gate{S1} & \gate{\shortmin S2} & \gate{QFT^\dagger} & \xymultigate{0.3em}{1.2em}{1}{\begin{tabular}[c]{@{}c@{}}$MAX$\\ $(0,\tilde{B})$ \end{tabular}} & \gate{QFT} & \qw 
& \push{=} && \gate{\mylim{B}{init}} & \xymultigate{.6em}{1.2em}{1}{B} & \qw\\
\lstick{\ket{a}}  & \qw  & \qw & \qw  & \qw  & \xyghost{0.3em}{1.2em}{\begin{tabular}[c]{@{}c@{}}$MAX$\\ $(0,\tilde{B})$ \end{tabular}} & \qw & \qw 
&&& \qw &\xyghost{.6em}{1.2em}{B} & \qw
}
\end{array}
\]
\caption{Calculation of buffer value $B_{out}$ for the simplified model after a single day}
    \label{fig:Boutoneday}
\end{circuit}

\subsection{Constraint checking} \label{subsec:constraintchecks}
After each day of operation, the buffer content must not exceed its maximum capacity $B_{max}$. We use this as a condition within Grover's algorithm, to distinguish feasible and infeasible solutions, by introducing a qubit $\ket{c_1}$ that will be $\ket{1}$ when the condition is met, and $\ket{0}$ otherwise. 

\begin{align}
c_1 &=
\begin{cases}
1 \textrm{ if} B_{out} \leq B_{max} \Rightarrow B_{out} - (B_{max}+1) < 0 \\
0 \textrm{ if } B_{out} > B_{max} 
\end{cases} \label{eq:caseforc1}
\end{align}


In order to set the value of $\ket{c_1}$, 
we first calculate $B_{out}-(B_{max}+1)$, and subsequently apply an inverse quantum Fourier transform, $QFT^\dagger$, to return the buffer qubit register to the computational basis. If the condition is met, then the buffer register state then corresponds to a negative (binary) number, and we can apply a CNOT from the MSB to our condition qubit $\ket{c_1}$. Finally, we need to restore the buffer register to its state before this constraint check, by applying a $QFT$ to return to the Fourier basis, and then adding $(B_{max}+1)$. The circuit for this operation is shown in \cref{fig:circuitforc1}, with the buffer register at the output again holding a value of $B_{out}$.


\begin{circuit}[hbt]
\[
\begin{array}{c}
\small
\Qcircuit @C=0.3em @R=0.5em {
&\lstick{\ket{q_{S1}}} & \qw & \ctrlxo{1} & \qw & \qw& \qw& \qw & \qw & \qw\\
&\lstick{\ket{q_{S2}}} & \qw & \ctrlxo{1} & \qw & \qw& \qw& \qw & \qw & \qw \\
&\lstick{\ket{q_{b}}} & \gate{\mylim{B}{init}} & \multigate{1}{B} & \gate{\shortmin(\mylim{B}{max}\!+\,1)} & \gate{QFT^\dagger} & \ctrl{2} & \gate{QFT} &  \gate{(\mylim{B}{max}\!+1)} & \qw \\
&\lstick{\ket{a}}  & \qw & \ghost{B} & \qw& \qw& \qw& \qw & \qw & \qw\\
&\lstick{\ket{c_1}}  & \qw & \qw & \qw& \qw& \targ&\rstick{\ket{1} \textrm{ if }B_{out} \leq B_{max}} \qw\\
}
\end{array}
\]
\caption{Evaluation of maximum buffer capacity condition $B_{out} \leq B_{max}$, with $B_{init}$ and $B$ as defined in \cref{fig:Boutoneday}. The CNOT is controlled by the MSB of the register $\ket{q_b}$. The buffer register is returned to its original state $B_{out}$ through a quantum Fourier transform and by adding the value of $B_{max}+1$.}
    \label{fig:circuitforc1}
\end{circuit}

The output volume $V_{out}$ of the simplified model is calculated as:
\begin{align*}
    V_{out} = B_{init}+S1-B_{final}
\end{align*}
where $B_{final}$ is the content of the buffer at the end of the day. The circuit will (initially) be applied only for 1 day, in which case $B_{final} = B_{out}$.
To avoid the additional gates required to calculate $-B_{final}$ (\cref{fig:negationcircuit}), 
we can instead calculate $-V_{out} = -B_{init}-S1+B_{final}$. We can calculate the output volume on the same qubits that are used to store the buffer occupancy value, which requires fewer qubits than using a separate qubit register and saves copying or recomputing the data in the buffer. This means that the volume requirement cannot easily be checked at the end of each day, but this is not required by the constraints of the simplified model. $-S1$ and $-B_{init}$ can be easily calculated using the Two's complement method as before. Because we are computing $-V_{out}$, the constraints for the output volume are defined accordingly. We call these output volume constraints $c_2$ and $c_3$, with values as in \cref{eq:caseforc2,eq:caseforc3}.

\begin{circuit}[htbp]
\[
\begin{array}{c}
\small
\Qcircuit @C=0.7em @R=0.7em {
\lstick{\ket{q_{S1}}} &\ctrlxo{2} & \ctrlxo{2} & \qw& \qw &&&
\ctrlxo{1} & \qw \\
\lstick{\ket{q_{S2}}} & \ctrlxo{1}& \qw & \qw & \qw & \push{\rule{0em}{0em}=\rule{0em}{0em}} &&
\ctrlxo{1} & \qw \\
\lstick{\ket{q_{b}}} & \gate{B_{final}} & \gate{\shortmin S1} & \gate{\shortmin B_{init}}& \qw &&&
\gate{\shortmin V_{out}} & \qw 
}
\end{array}
\]
\caption{$-V_{out} = -B_{init}-S1+B_{final}$}
    \label{fig:Vout}
\end{circuit}


\begin{align}
c_2 &=
\begin{cases}
1 \textrm{ if } V_{out} \geq V_{low} \Rightarrow -V_{out} + V^* - \Delta \leq 0 \\
0 \textrm{ if } V_{out} < V_{low} \Rightarrow -V_{out} > -V_{low} 
\end{cases} \label{eq:caseforc2} \\
c_3 &=
\begin{cases}
1 \textrm{ if } V_{out} \leq V_{up} \Rightarrow -V_{out} + V^* + \Delta \geq 0 \\
0 \textrm{ if } V_{out} > V_{up} \Rightarrow -V_{out} < -V_{low}   \label{eq:caseforc3} 
\end{cases}
\end{align}
The lower limit for the tolerance is $V_{low} = V^* - \Delta$, for target volume $V^*$  and tolerance $\Delta$. The upper limit for the tolerance is defined as $V_{up} = V^* + \Delta$.

To set qubit $c_2$, we need to transform the inequality from \cref{eq:caseforc2} to $-V_{out} + (V^* - \Delta - 1) < 0$ (where '1' is the minimum difference between numbers), which can be calculated on the buffer register. Then we use a CNOT from the MSB of the buffer register to qubit $c_2$ to flip the qubit to $\ket{1}$ when the condition is met, and keep it at $\ket{0}$ otherwise.

For qubit $c_3$, we can do the same. We calculate $-V_{out} + (V^* + \Delta)$ and then set $c_3$ based on the MSB of the result. An X-gate is needed to flip $c_3$, since the condition is met when the MSB is in state $\ket{0}$ (when the result of the calculation is bigger than or equal to zero).

To avoid recomputing $V_{out}$, we can compute the result for evaluating $c_3$ from the result for $c_2$, simply by adding $(2\Delta + 1)$ to the latter. This works because we are first comparing to the lower limit, and the output volume should be within $2\Delta$ of it to be within the upper limit. The addition of '1' is again the minimum precision.

Combining these will give the circuit in \cref{fig:alternative26}.
\begin{circuit*}[hbtp]
\[
\begin{array}{c}
\small
\Qcircuit @C=0.5em @R=.5em {
\lstick{\ket{q_b}} & \gate{\shortmin V_{out}} & \gate{V^*-\Delta-1} & \gate{QFT^\dagger} & \ctrl{1} & \gate{QFT} & \gate{2\Delta + 1} & \gate{QFT^\dagger} & \ctrl{2} & \gate{QFT} & \qw \\
\lstick{\ket{c_2}=\ket{0}} & \qw             & \qw               & \qw                & \targ    & \qw        & \qw                & \qw                & \qw &  \rstick{\ket{1} \textrm{ if }V_{out} \geq V_{low}} \qw \\
\lstick{\ket{c_3}=\ket{0}} & \qw             & \qw               & \qw                & \qw      & \qw        & \qw                & \gate{X}           & \targ  & \rstick{\ket{1} \textrm{ if }V_{out} \leq V_{up}} \qw
}\end{array}\]
\caption{Circuit for evaluating the output volume conditions $\ket{c_2}$ and $\ket{c_3}$.}
    \label{fig:alternative26}
\end{circuit*}

\subsection{Completed circuit for day one} \label{subsec:completedcircuit} 
The complete circuit for scheduling one day is shown in \cref{fig:completecircuit}. This circuit can be used as the oracle in the circuit in \cref{fig:groversoracle}, where the sub-circuit within the dashed box is applied a specific number of times, namely the number of Grover rotations. The diffuser operator in \cref{fig:groversoracle} is implemented as \cref{fig:diffuser}.

\begin{circuit*}[hbtp]
\[
\begin{array}{c}
\small
\Qcircuit @C=1em @R=0.4em {
\lstick{\ket{q_{S1}}} & / \qw & \gate{H} & \qw & \ctrlxo{2} 
& \ar@{--}[]+<1.7em,1em>;[ddddddd]+<1.7em,-1em>  \qw 
& \ar@{--}[]+<2.5em,1em>;[ddddddd]+<2.5em,-1em> \qw 
&\ar@{--}[]+<2.7em,1em>;[ddddddd]+<2.7em,-1em>  \qw
& \ar@{--}[]+<0.8em,1em>;[ddddddd]+<0.8em,-1em> \qw  & \qw & \ctrlxo{2} & \qw 
& \ar@{--}[]+<2.2em,1em>;[ddddddd]+<2.2em,-1em> \qw 
& \ar@{--}[]+<0.7em,1em>;[ddddddd]+<0.7em,-1em> \qw
& \ar@{--}[]+<2em,1em>;[ddddddd]+<2em,-1em> \qw& \qw& \qw & \qw\\
\lstick{\ket{q_{S2}}} & / \qw & \gate{H}        & \qw & \qw & \ctrlxo{1} & \qw & \qw& \qw& \qw & \qw & \qw & \qw& \qw& \qw& \qw & \qw & \qw\\
\lstick{\ket{q_{b}}}  & / \qw & \gate{H}         & \gate{\mylim{B}{init}} & \gate{S1} & \gate{\shortmin S2} & \multigate{1}{\parbox{3em}{$MAX$\\ $(0,\tilde{B})$}} & \gate{\shortmin(\mylim{B+1}{max})} & \ctrl{2} & \gate{\mylim{B+1}{max}}  & \gate{\shortmin S1} & \gate{\shortmin\mylim{B}{init}} &  \gate{\mylim{V-1}{low}}& \ctrl{3}& \gate{2\Delta+1}& \ctrl{4} & \qw & \qw\\
\lstick{\ket{a}}      & / \qw & \qw              & \qw  & \qw   & \qw & \ghost{\parbox{3em}{$MAX$}}& \qw & \qw& \qw & \qw & \qw& \qw& \qw& \qw & \qw & \qw & \qw\\
\lstick{\ket{c_1}}  -  &   \qw & \qw              &\qw & \qw & \qw & \qw& \qw& \targ& \qw & \qw & \qw& \qw& \qw& \qw & \qw & \ctrl{1} & \qw \\
\lstick{\ket{c_2}}  -  &   \qw & \qw              &\qw & \qw & \qw & \qw& \qw& \qw& \qw & \qw & \qw& \qw& \targ& \qw & \qw & \ctrl{1} & \qw \\
\lstick{\ket{c_3}}  -  &   \qw & \qw              &\qw & \qw & \qw & \qw& \qw& \qw& \qw & \qw & \qw& \qw& \qw& \qw & \targ & \ctrl{1} & \qw \\
\lstick{\ket{-}}  -  &  \qw & \qw &  \qw \qw  & \qw & \qw& \qw& \qw& \qw & \qw & \qw& \qw& \qw& \qw & \qw & \qw & \targ & \qw
}
\end{array}
\]
\caption{The complete circuit for the simplified model with buffer and volume constraints for scheduling of a single day. The $QFT^\dagger$ and $QFT$ operations have been abstracted as vertical dashed lines. Every time a dashed line is crossed, the buffer is converted from the Fourier basis to the computational basis or back. This circuit is used as the oracle in Grover's search algorithm.}
    \label{fig:completecircuit}
\end{circuit*}

\begin{circuit}[htbp]
\[
\begin{array}{c}
\small
\Qcircuit @C=0.7em @R=.7em {
&\push{\ket{q_{S1,1}}~} & \multigate{3}{\mathit{Diffuser}} & \qw & && \gate{H} & \gate{X} &  \ctrl{1} & \gate{X} & \gate{H} & \qw  \\
&\push{\ket{q_{S1,2}}~} & \ghost{\mathit{Diffuser}} & \qw & && \gate{H} & \gate{X} &  \ctrl{1} & \gate{X} & \gate{H} & \qw  \\
&\push{\ket{q_{S2,1}}~} & \ghost{\mathit{Diffuser}} & \qw &\ustick{=} \push{\rule{0em}{0em}\rule{0em}{0em}} && \gate{H} & \gate{X} &  \ctrl{1} & \gate{X} & \gate{H} & \qw  \\
&\push{\ket{q_{S2,2}}~} & \ghost{\mathit{Diffuser}} & \qw &&& \gate{H} & \gate{X} &  \gate{Z} & \gate{X} & \gate{H} & \qw  \\
}
\end{array}
\]
\caption{Diffuser for Grover's algorithm}
    \label{fig:diffuser}
\end{circuit}

All possible combinations of shifts for one day are shown in \cref{tab:expectedoutputSSSM}. For each possibility, the buffer content and output volume have been calculated, as well as the expected value of the condition checks.

The schedules where shop 1 has a long shift and shop 2 has a short shift are not valid, because, at the end of the day there will be more units in the buffer than its maximum capacity. The target volume of $8\pm 0.4$ units per day cannot be met when scheduling for a single day, since shop 2 can only produce 7 or 9 units in a shift. 
If we relax the constraint (set $\Delta$ to 1), the production volume never exceeds the maximum. The minimum production target is met if shop 2 has a shift of at least 7 hours. 
There are 16 possible schedules for the simplified model, of which six meet all the requirements.

\begin{table}
\caption{Expected output for the simplified model for one day, bolded rows are solutions that satisfy all the constraints ($c_1$: $B_{out} \leq B_{max}$, $c_2$: $V_{out} \geq V_{low}$, $c_3$: $V_{out} \leq V_{up}$)}
\label{tab:expectedoutputSSSM}
\centering
\setlength{\tabcolsep}{3pt}
\begin{tabular}{rclllllllll}
\multicolumn{2}{c|}{$\ket{S1}~\ket{S2}$}       & \multicolumn{1}{l|}{$B_{init}$} & $U(S1)$     & \multicolumn{1}{l|}{$U(S2)$}    & $B_{out}$ & $V_{out}$ & \multicolumn{1}{l|}{\textit{Cost}} & $c_1$ & $c_2$ & $c_3$ \\ \midrule
00          & \multicolumn{1}{c|}{00}          & \multicolumn{1}{l|}{5}          & 0           & \multicolumn{1}{l|}{0}          & 5         & 0 & \multicolumn{1}{l|}{\$0}          & 1               & 0               & 1              \\
00          & \multicolumn{1}{c|}{01}          & \multicolumn{1}{l|}{5}          & 0           & \multicolumn{1}{l|}{4}          & 1         & 4 & \multicolumn{1}{l|}{\$4}          & 1               & 0               & 1              \\
00          & \multicolumn{1}{c|}{10}          & \multicolumn{1}{l|}{5}          & 0           & \multicolumn{1}{l|}{7}          & 0         & 5 & \multicolumn{1}{l|}{\$7}          & 1               & 0               & 1              \\
00          & \multicolumn{1}{c|}{11}          & \multicolumn{1}{l|}{5}          & 0           & \multicolumn{1}{l|}{9}          & 0         & 5 & \multicolumn{1}{l|}{\$9}          & 1               & 0               & 1              \\ \midrule
01          & \multicolumn{1}{c|}{00}          & \multicolumn{1}{l|}{5}          & 5           & \multicolumn{1}{l|}{0}          & 10        & 0 & \multicolumn{1}{l|}{\$5}          & 1               & 0               & 1              \\
01          & \multicolumn{1}{c|}{01}          & \multicolumn{1}{l|}{5}          & 5           & \multicolumn{1}{l|}{4}          & 6         & 4 & \multicolumn{1}{l|}{\$9}          & 1               & 0               & 1              \\
\textbf{01} & \multicolumn{1}{c|}{\textbf{10}} & \multicolumn{1}{l|}{\textbf{5}} & \textbf{5}  & \multicolumn{1}{l|}{\textbf{7}} & \textbf{3} & \textbf{7} & \multicolumn{1}{l|}{\textbf{\$12}} & \textbf{1}      & \textbf{1}      & \textbf{1}     \\
\textbf{01} & \multicolumn{1}{c|}{\textbf{11}} & \multicolumn{1}{l|}{\textbf{5}} & \textbf{5}  & \multicolumn{1}{l|}{\textbf{9}} & \textbf{1} & \textbf{9} & \multicolumn{1}{l|}{\textbf{\$14}} & \textbf{1}      & \textbf{1}      & \textbf{1}     \\ \midrule
10          & \multicolumn{1}{c|}{00}          & \multicolumn{1}{l|}{5}          & 8           & \multicolumn{1}{l|}{0}          & 13         & 0 & \multicolumn{1}{l|}{\$8}          & 0               & 0               & 1              \\
10          & \multicolumn{1}{c|}{01}          & \multicolumn{1}{l|}{5}          & 8           & \multicolumn{1}{l|}{4}          & 9          & 4 & \multicolumn{1}{l|}{\$12}          & 1               & 0               & 1              \\
\textbf{10} & \multicolumn{1}{c|}{\textbf{10}} & \multicolumn{1}{l|}{\textbf{5}} & \textbf{8}  & \multicolumn{1}{l|}{\textbf{7}} & \textbf{6} & \textbf{7} & \multicolumn{1}{l|}{\textbf{\$15}} & \textbf{1}      & \textbf{1}      & \textbf{1}     \\
\textbf{10} & \multicolumn{1}{c|}{\textbf{11}} & \multicolumn{1}{l|}{\textbf{5}} & \textbf{8}  & \multicolumn{1}{l|}{\textbf{9}} & \textbf{4} & \textbf{9} & \multicolumn{1}{l|}{\textbf{\$17}} & \textbf{1}      & \textbf{1}      & \textbf{1}     \\ \midrule
11          & \multicolumn{1}{c|}{00}          & \multicolumn{1}{l|}{5}          & 10          & \multicolumn{1}{l|}{0}          & 15         & 0 & \multicolumn{1}{l|}{\$10}          & 0               & 0               & 1              \\
11          & \multicolumn{1}{c|}{01}          & \multicolumn{1}{l|}{5}          & 10          & \multicolumn{1}{l|}{4}          & 11         & 4 & \multicolumn{1}{l|}{\$14}          & 0               & 0               & 1              \\
\textbf{11} & \multicolumn{1}{c|}{\textbf{10}} & \multicolumn{1}{l|}{\textbf{5}} & \textbf{10} & \multicolumn{1}{l|}{\textbf{7}} & \textbf{8} & \textbf{7} & \multicolumn{1}{l|}{\textbf{\$17}} & \textbf{1}      & \textbf{1}      & \textbf{1}     \\
\textbf{11} & \multicolumn{1}{c|}{\textbf{11}} & \multicolumn{1}{l|}{\textbf{5}} & \textbf{10} & \multicolumn{1}{l|}{\textbf{9}} & \textbf{6} & \textbf{9} & \multicolumn{1}{l|}{\textbf{\$19}} & \textbf{1}      & \textbf{1}      & \textbf{1}   \\ \bottomrule
\end{tabular}
\end{table}

\begin{figure}[tbp]
    \centering
    \includegraphics[width=\linewidth]{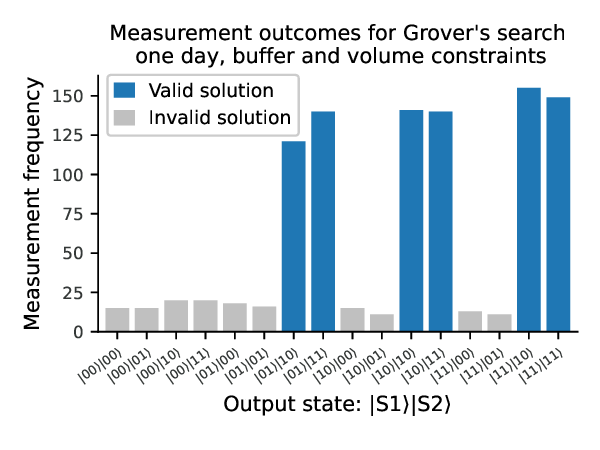}
    \caption{Simulated measurement outcomes using Qiskit for the implementation of Grover's search for scheduling one day and only buffer and volume constraints. States corresponding to valid solutions are shown in dark blue.}
    \label{fig:groversoutput}
\end{figure}

To validate that our approach gives the expected results,\cref{fig:completecircuit} was implemented in Qiskit and simulated with Qiskit Aer, assuming noiseless, fully-connected qubits. The results from 1000 measurements of the output state of the circuit are shown in \cref{fig:groversoutput}, with the six valid solutions clearly amplified.

\subsection{Extending to multiple days} \label{subsec:extending}


To extend the oracle to multiple days, we need to implement several changes compared to the circuit for a single day. Firstly, to encode the shifts for each shop for all possible days, we will need $2n$ qubits per shop for $n$ days.

Secondly, we need to calculate the buffer occupancy values and check the buffer constraints after each day. The buffer value after one day is $B_1 = \mathit{MAX}(0,\tilde{B}_1)$, where $\tilde{B}_1 = B_{init} + S1_1 - S2_1$. Combined, this gives $B_1 = \mathit{MAX}(0, B_{init} + S1_1-S2_1)$. The buffer value after two days will be $B_2 = \mathit{MAX}(0,B_1 + S1_2 - S2_2)$, and after $n$ days it will be:
\begin{align*}
    B_n = \mathit{MAX}(0,B_{n-1} + S1_n-S2_n)
\end{align*}
For each day, the values produced by each shop are added or subtracted from the buffer register, controlled by the qubits in the corresponding shop register. Then the value for $MAX(0,\tilde{B})$ need to be calculated, which requires a new set of ancilla qubits ($\ket{a_n}$) per day to keep the circuit reversible. Finally, a conditional qubit ($\ket{c_{1n}}$) is set depending on if the buffer exceeds the maximum allowed buffer content. 

This means that we repeat everything in \cref{fig:Boutoneday} after the gate for $B_{init}$, with separate controls and ancillas for each day. For two days, this looks like the circuit in \cref{fig:Bouttwodays}. After $n$ repeats for $n$ days, the output volume can be calculated. This can be done using the same circuit as for one day (\cref{fig:Vout}), but instead of subtracting the production of S1 once, it is done $n$ times for $n$ days. 

\begin{circuit*}[htbp]
\[
\begin{array}{c}
\small
\Qcircuit @C=0.5em @R=0.5em {
\lstick{\ket{q_{S1,1}}}  & \ctrlxo{4} & \ustick{\textrm{\textbf{\normalsize Day~1}}} \qw & \qw & \qw& \qw & \qw  &  \qw & \qw & \qw &  \ustick{\textrm{\textbf{\normalsize Day~2}}} \qw & \qw & \qw & \qw & \qw& \qw & \qw  & \qw \\
\lstick{\ket{q_{S1,2}}} & \qw & \qw & \qw & \qw& \qw & \qw  & \qw & \qw & \ctrlxo{3} & \qw \qw & \qw & \qw & \qw & \qw& \qw & \qw  & \qw \\
\lstick{\ket{q_{S2,1}}} & \qw & \ctrlxo{2} & \qw & \qw& \qw & \qw  & \qw & \qw & \qw & \qw & \qw & \qw & \qw & \qw& \qw & \qw  & \qw \\
\lstick{\ket{q_{S2,2}}}  & \qw & \qw & \qw & \qw& \qw & \qw  & \qw & \qw & \qw & \ctrlxo{1} & \qw & \qw & \qw & \qw& \qw & \qw  & \qw \\
\lstick{\ket{q_{b}}}  & \gate{S1} & \gate{\shortmin S2} & \gate{QFT^\dagger} & \multigate{1}{\parbox{2.5em}{$MAX$\\ $(0,\tilde{B})$}} & \gate{QFT} & 
\gate{\shortmin(\mylim{B}{max}\!\!+1)} & \ctrl{3} & \gate{\mylim{B}{max}\!\!+1}
& \gate{S1} & \gate{\shortmin S2} & \gate{QFT^\dagger} & \multigate{2}{\parbox{2.5em}{$MAX$\\ $(0,\tilde{B})$}} & \gate{QFT} & 
\gate{\shortmin(\mylim{B}{max}\!\!+1)} & \ctrl{4} & \gate{\mylim{B}{max}\!\!+1}  & \qw \\
\lstick{\ket{a1}}   & \qw & \qw  & \qw  & \ghost{\parbox{2.5em}{$MAX$}} & \qw & \qw  & \qw & \qw & \qw & \qw & \qw & \nghost{\parbox{2.5em}{$MAX$}} \qwdash & \qwdash &\qw & \qw & \qw  & \qw \\
\lstick{\ket{a2}}   & \qw & \qw  & \qw  & \qw & \qw & \qw  & \qw & \qw & \qw & \qw & \qw & \ghost{\parbox{2.5em}{$MAX$}} & \qw & \qw& \qw & \qw  & \qw \\
\lstick{\ket{c_{1,1}}} & \qw & \qw & \qw & \qw& \qw & \qw  & \targ & \qw & \qw& \qw & \qw & \qw & \qw & \qw& \qw & \qw  & \qw \\
\lstick{\ket{c_{1,2}}} & \qw & \qw & \qw & \qw& \qw & \qw  & \qw & \qw & \qw & \qw & \qw & \qw & \qw & \qw& \targ & \qw  & \qw 
{\POS"1,2"."5,2"."5,9"."8,8"!C*+<0.5em>\frm{--}} {\POS"2,10"."5,10"."5,17"."9,16"!C*+<0.5em>\frm{--}}\\
}
\end{array}
\]
\caption{Calculation of buffer value $B_{out}$ for the simplified model for 2 days, including calculation of $MAX(0,\tilde{B})$ for each day. Conditional qubits $\ket{c_{1,1}}$ and $\ket{c_{1,2}}$ are set depending on if the buffer exceeds the maximum allowed buffer content for both days.}
    \label{fig:Bouttwodays}
\end{circuit*}


Checking whether the output volume falls within the allowed limits is similar to the case of a single day, since it only needs to be checked at the end. The only differences are the target output volume $V^*$ and the allowable margin $\Delta$, which need to be adjusted depending on the number of days that are being simulated.

\begin{figure}[htbp]
    \centering
    \includegraphics[width=\linewidth]{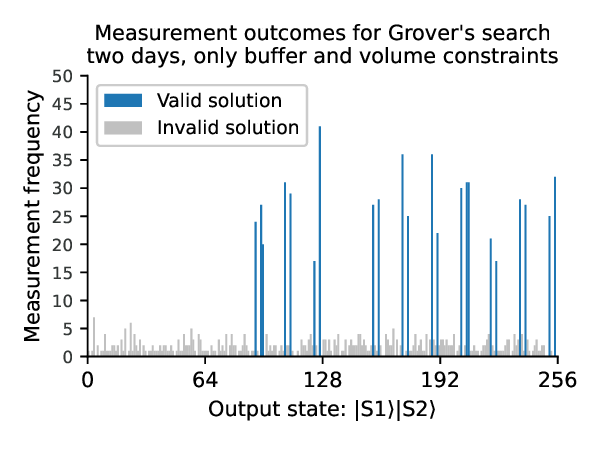}
    \caption{Simulated measurement outcomes using Qiskit for the implementation of Grover's search for scheduling two days and only buffer and volume constraints. States corresponding to valid solutions are shown in dark blue, output states are converted from binary to decimal.}
    \label{fig:groversoutput2days}
\end{figure}

For verification, we implemented \cref{fig:Bouttwodays} (corresponding to two days of operation) in Qiskit and simulated 1000 circuit shots using Qiskit Aer, which produced the measurement results shown in \cref{fig:groversoutput2days}. The 22 valid solutions (shown in blue) are clearly amplified over the 244 non-valid solutions.

\subsection{Cost constraint} \label{sec:costconstraint}

In the framework of Grover adaptive search, the objective function corresponding to the total factory operating cost can be handled as an additional constraint by setting a certain maximum allowed cost. To store this total cost, we introduce an additional qubit register, which must be large enough to avoid numerical overflow when both shops are operated for the maximum number of hours daily. We also add a condition qubit for the cost constraint.

Computing the total cost is straightforward: we add the cost corresponding to each shift to the register with the same controlled adders as we did for the shifts. The same set of shop qubits also controls them. In the case of the simplified model, the cost and the number of produced units are the same, but this is not necessarily the case for more realistic models. 

Adding the cost can be done at any point in the circuit, but for simplicity's sake, we will add the costs for each shop just after the corresponding values have been added to the buffer. 

At the end of the circuit, the cost constraint is checked with a similar construction as for the maximum and minimum volume constraints. To avoid adding the minimum precision, we define the cost constraint as $C_{out} < C_{max}$ where $C_{out}$ is the calculated cost of the solution, and $C_{max}$ is the maximum allowed cost. This means that we can calculate  $C_{out} - C_{max}$ at the end of the circuit, and the condition is met if the result is less than zero, which means that the MSB is 1. We use this to set a single condition qubit with a CNOT. 

\begin{circuit}[htb]
\[
\begin{array}{c}
\small
\Qcircuit @C=0.4em @R=0.5em{
&\lstick{\ket{q_{S1,1}}} & \ctrlxo{4} & \qw        & \ctrlxo{6} & \qw        & \qw & \qw & \qw & \qw        & \qw & \qw & \qw \\
&\lstick{\ket{q_{S1,2}}} & \qw        & \ctrlxo{3} & \qw        & \ctrlxo{5} & \qw & \qw & \qw & \qw        & \qw & \qw & \qw \\
&\lstick{\ket{q_{S2,1}}} & \qw        & \qw        & \qw        & \qw        & \qw & \qw & \ctrlxo{2} & \qw & \ctrlxo{4} & \qw & \qw \\
&\lstick{\ket{q_{S2,2}}} & \qw        & \qw        & \qw        & \qw        & \qw & \qw & \qw & \ctrlxo{1} & \qw & \ctrlxo{3} & \qw \\
&\lstick{\ket{q_b}}      & \gate{S1}  & \gate{\shortmin S2} & \qw        & \qw        & \qw & \multigate{1}{\parbox{2.5em}{Buffer\\checks}} & \gate{S1} & \gate{\shortmin S2} & \qw  & \qw  & \qw \\
&\lstick{\ket{a}\!+\!\ket{c_1}} & \qw        & \qw        & \qw        & \qw        & \qw &  \ghost{\parbox{2.5em}{Buffer}}  & \qw& \qw & \qw & \qw & \qw \\
&\lstick{\ket{q_\$}}     & \qw        & \qw        & \gate{C1}  & \gate{C2}  & \qw & \qw & \qw & \qw & \gate{C1} & \gate{C2} & \qw \\
}
\end{array}
\]
\caption{Calculation of the buffer content and the cost for shops $S1$ and $S2$ for two days. All elements are as defined before, with additionally cost register $\ket{q_\$}$, and gates $C1$ and $C2$, which are the costs per shift for shops 1 and 2.} \label{fig:circuitwithcosts}
\end{circuit}

\begin{figure*}[hbt]
    \centering
        \includegraphics[width=\linewidth]{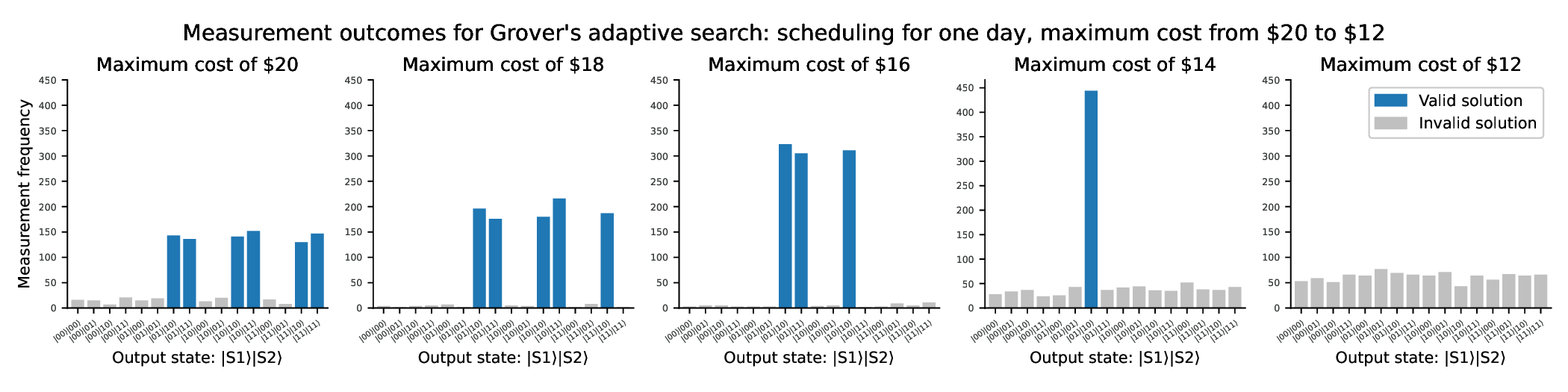}
    \caption{Simulated measurement outcomes for Grover's adaptive search for scheduling 1 day with manually set value for the cost constraint ($C_{out} < C_{max}$) starting at \$20 and decreasing with steps of \$2 until \$12, at which point there are no solutions left that satisfy all the constraints.}
    \label{fig:GAS}
\end{figure*}

\subsection{Validating the stop condition of GAS} \label{GASstopcondition} 
Having constructed the oracle and verified its ability to produce expected solutions, we now proceed to investigate a proper \textit{stop condition} for (see \cref{alg:GAS}).

There are many options for the stop condition of GAS: the loop can be stopped when no improvement has been achieved in the last few of iterations when a threshold lowest cost has been achieved, when a certain amount of time has passed, when a certain number of iterations has been performed, etc. 

We will stop the GAS procedure when the combined total rotation count (number of times that Grover's rotation operator has been applied) exceeds $\sqrt{N}$, where $N$ is the size of the search space. To validate our decision, we have simulated 
10,000 runs of GAS and kept track of the number of rotations and the current best cost after each loop. The number of valid solutions and associated costs correspond to those for the shift schedules for three subsequent days. Each shop has four possible shift lengths per day, so the total number of possible schedules $N =4^{2n}$. For three days, the total problem size is thus $N = 4^{2\cdot3} = 4096$.

\begin{figure}[htbp]
    \centering
    \includegraphics[width=\linewidth]{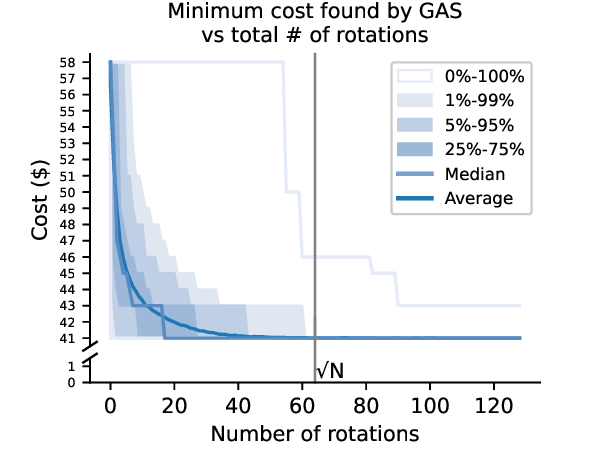}
    \caption{Minimum cost found by GAS after a certain number of rotations for finding shift schedules for three days, with the average, median and percentile intervals calculated from 10,000 runs of the adaptive search.}
    \label{fig:min_cost_GAS_illustrate_3days}
\end{figure}

Each simulated run of GAS ran for a total of $2\sqrt{N} = 128$ rotations. For each loop of QISS, the specific number of rotations was determined randomly as outlined in \cref{alg:GAS}. The initial cost was set to \$60, which is three dollars more than the cost of operating both shops, the maximum allowed number of hours for all three days. When a valid solution is found, i.e. a solution that satisfies all the constraints, the initial cost is replaced with the cost for the valid solution. This allows us to see how many rotations it took QISS to arrive at a valid solution. After each loop was completed, the running totals for the number of rotations with corresponding lowest found cost were stored. After 10,000 runs of GAS, the average, median, and percentile intervals were calculated from the results. These can be found in \cref{fig:min_cost_GAS_illustrate_3days}.

As can be seen from the figure, the minimum cost found by QISS decreases exponentially with the number of Grover rotations, and the chance that we have landed on the cheapest possible shift schedule increases accordingly. After a total of $\frac{\pi}{4}\sqrt{N} = 51$ rotations, more than 95\% of runs have landed on the minimum cost of 41. After $\sqrt{N}$ rotations, this increases to more than 99\%. Therefore, we will set the stop condition of GAS to a number of rotations equal to $\sqrt{N}$, which yields good results while still ensuring an asymptotic quadratic speedup over a classical brute-force search.

The exact rate at which the found cost decreases is dependent on the problem size, how many possible solutions are valid, the distribution of the costs associated with each valid solution, and the number of valid solutions that share the minimum cost. We do not have this data for larger problem sizes, but we expect that these results hold up for larger problem sizes.

\section{Conclusion and future work} \label{sec:conclusion}

We have introduced QISS, a quantum algorithm based on Grover's adaptive search for industrial shift scheduling problems with production target and intermediary storage constraints, a situation found in settings such as the automotive industry. We show the construction of a quantum circuit to implement the necessary Grover's oracle for an arbitrary number of days of factory operations, within the context of a simplified model comprising two shops and one buffer. For small problem instances, we have numerically corroborated the performance of the algorithm. In particular, in our examples we verify that $\sqrt{N}$ applications of Grover’s rotation operator suffice to find the optimal solution, where $N$ is the size of the solution space. This shows that an asymptotic quadratic speedup can in principle be achieved over classical unstructured search. In practice, this may be useful in at least two scenarios: (1) to obtain exact solutions to problems of small or modest size, in the context of benchmarking heuristic algorithms designed to scale to much larger, industrial problem sizes; (2) to use QISS as an integral component of a heuristic strategy, where exact solutions for short time periods are used to construct a solution for a longer time period.

Our work lays the foundation for future research in several directions. Firstly, in the context of the simplified model, QISS could be used as a target algorithm to study and benchmark the performance of quantum computing systems. Specifically, in the pre-fault-tolerant era one could test the different circuit primitives, and investigate their susceptibility to noise on different quantum computing platforms. Techniques to suppress, mitigate, or detect certain types of error may be found, which could allow for improved results, and yield useful insights for running the algorithm in future on fully fault-tolerant machines. In the fault-tolerant era, QISS could be incorporated into application-level benchmark frameworks~\cite{Finzgar_2022}.

Secondly, we have focused on a simplified shift scheduling model with only two shops and a single buffer. It would be interesting to develop Grover's oracles for more complex, more realistic instances of industrial shift scheduling, such as the one shown in Figure 1. For example, new circuit primitives may need to be designed due to the intricacies resulting from multiple and/or shared buffers (see Appendix). Moreover, while the resulting circuits would very quickly become intractable to simulate on a classical computer, even for a single day, it would allow a quantitative investigation of the additional circuit complexity arising from the presence of multiple shops and buffers.

Thirdly, in this work we have developed quantum circuits at the \textit{logical} level, which assumes qubits to be perfectly noiseless. To implement these circuits in practice on quantum computers, which are always subject to some degree of noise, a fault-tolerant quantum error correction scheme would be necessary. Quantum error correction introduces significant resource overheads in terms of the number of required qubits and gate operations. Key quantities of interest such as the total number of \textit{physical} qubits and the total computational runtime can be estimated through frameworks incorporating different layers of assumptions on the architecture of a fault-tolerant quantum computer, and the compilation of quantum programs \cite{art:resourceestpaper, misc:qualtran}. Subsequently, one could compare these estimates to those of classical computing approaches, and seek to identify the scale at which QISS may deliver a speedup in practice for unstructured search, taking into account the gap in execution time for basic quantum and classical circuit operations \cite{art:GoogleBeyondQuadratic, art:TroyerBeyondQuadratic}.

Finally, we have not investigated whether the industrial shift scheduling problem is amenable to solution by dynamic programming methods, however we note that quantum algorithms capable of exploiting optimal substructure have been described in \cite{art:AmbainisQuantumDynamicProgramming}. Therefore, if a speedup over unstructured search can be obtained classically with dynamic programming, it may also be possible to incorporate this into QISS.

\section*{Code availability}
The code to generate the figures in this paper can be found at \href{https://github.com/anneriet/QISS}{https://github.com/anneriet/QISS}. The full implementation of QISS is located in the \mbox{“QISS/QISS\_with\_cost.ipynb“} Python notebook and additionally in the {“QISS/operator\_definitions.py“} python file.

\section*{Acknowledgements}
We thank Vladislav Samoilov and Nikolas Beulich for providing the initial problem statement and for stimulating discussions.

\printbibliography

\appendix \label{the_appendix}

\section{Analysis of simplified model}

In this appendix, we provide supplementary information about the simplified model for shift scheduling introduced in \cref{sec:sssm}. While none of the details given here are necessary for the construction of the QISS algorithm presented in the main text, they do provide useful context on the industrial shift scheduling problem itself.

\subsection{Structure of solution space}\label{app:sol_space_structure}

To discover methods that more efficiently explore the solution space than a pure brute-force search, an important question is whether feasible solutions (those that satisfy all the constraints) exhibit any discernible structure. If they do, then one can potentially (partially) exclude infeasible regions from the search. 

In the industrial shift scheduling problem, we note that the output production volume constraint can in principle be used in this way. We have both a lower and upper total number of units that can be produced, i.\,e., $V_- = 8n - \Delta$ and $V_+ = 8n+\Delta$, respectively. We can easily convert these into the required number of hours that must be worked in total by each of the two shops. For example, to achieve an output of $V_-$, at a minimum, shop 1 must work $V_- - B_{init}$ hours, and shop 2 must work $V_-$ hours. A similar condition can be derived to ensure that no more than $V_+$ units are produced. 

These conditions on the shop working hours could then be used to avoid searching through solutions where the production volume is outside the allowed range. For instance, a classical brute force algorithm could be restricted in its scope by only checking solutions satisfying the above conditions. In the case of Grover's algorithm, in principle, one could prepare (e.\,g., using a quantum random access memory \cite{art:qram_Lloyd}) the initial state as a superposition only of solutions that respect the conditions. While the overall complexity of the problem is unchanged, in practice, the restriction to a subspace could lead to a significant speedup over a full brute-force search. Nevertheless, an exhaustive search within the identified subspace would still be necessary to obtain an exact solution. \ewannote{We have not investigated whether the industrial shift scheduling problem is amenable to solution by dynamic programming methods, however we note that quantum algorithms capable of exploiting optimal substructure have been described in Ref. \cite{art:AmbainisQuantumDynamicProgramming}. Therefore, if a speedup over unstructured search can be obtained classically with dynamic programming, it may also be possible to incorporate this into QISS.}
\annenote{This might also be interesting to note in the future work?}

\begin{figure}[t]
    \centering
    \includegraphics[width=\linewidth]{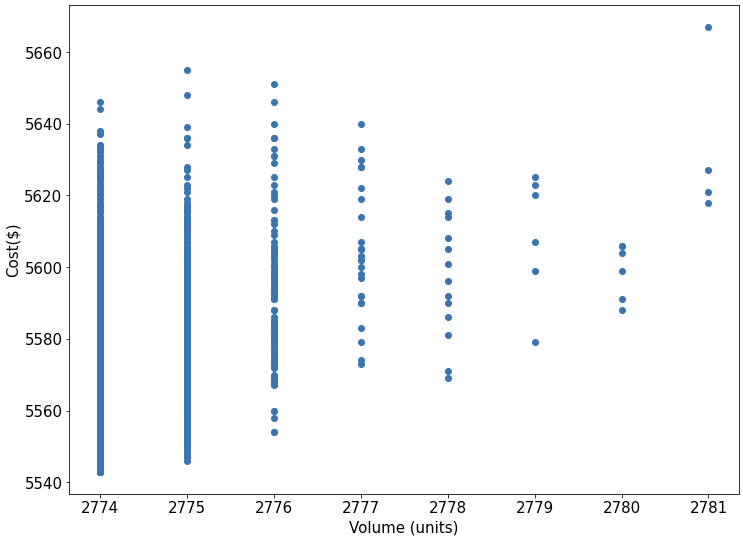}
    \caption{Total cost vs. output volume for 1623 feasible solutions (each represented by a blue dot) of the $n=365$ days simplified shift scheduling model described in \cref{sec:sssm}. The feasible solutions were obtained using simulated annealing.}
    \label{fig:cost_vol_plot}
\end{figure}

To illustrate the nature of the feasible solution space for the simplified model for the case of $n=365$ days, \cref{fig:cost_vol_plot} shows the relationship between the total cost and the total output volume for a total of 1623 feasible solutions found using a simulated annealing algorithm. A notable feature is the large variance in the cost of solutions that achieve the same output volume. For example, of the solutions achieving an output of 2774 units, there is a difference of around \$100 in the cost of the solutions found. This shows that cost minimization is not necessarily achieved simply by limiting production.

\subsection{A lower bound for the cost}

A simple way to compute a lower bound on the cost for the simplified model follows directly from considering the volume target. At the very least, we know that any feasible solution must result in an output of $V^* -\Delta$ vehicles. Therefore, we know with certainty that we must incur a minimum operating cost that corresponds to the labor required to produce those vehicles. 

Denoting our lower bound for the cost value as $C_{LB}$, we can compute it simply as the sum of the cost of the work required by both shops:   

\begin{equation}
    C_{LB} = \left(V^* - \Delta - B_{init}\right) + \left(V^* - \Delta\right)
\end{equation}

For the data specified in \cref{sec:sssm}, we find that ${C_{LB} = \$5543}$.

Such a formula is easily generalizable to the case of multiple shops and buffers, multiple vehicle types, and different hourly costs and hourly production rates for the shops. It may be possible to find a tighter lower bound than this, a question that we leave for future work.

Note that this method of obtaining a lower bound does not yield an explicit solution (shift configuration) corresponding to the cost $C_{LB}$. Moreover, we emphasize that $C_{LB}$ is not necessarily the cost of the (unknown) optimal solution, which we denote $C_{OPT}$, although it is certainly the case that $C_{LB} \leq C_{OPT}$. Coincidentally, as we see from \cref{fig:cost_vol_plot}, for the simplified model defined in \cref{sec:sssm}, we do indeed have $C_{OPT} = C_{LB}$. However, this may not hold if the data specifying the model were to be modified, and it is also unlikely to hold for more complex versions of the industrial shift scheduling problem.

\subsection{Beyond the simplified model}

More complicated versions of the industrial shift scheduling problem can consist of multiple shops, multiple buffers, multiple vehicle types, and non-linear production line structures, as depicted in \cref{fig:FSM}. One complication in tackling these problems is the increased size of the solution space, but there are a number of additional issues that can complicate the structure of the solution space itself.


For a given shift configuration, for any given shift, each shop is scheduled to work a certain number of hours, which in turn determines how many units can be processed during that shift. In the simplified model, the shop can be idle if the upstream (i.e. preceding) buffer is emptied during a shift, but a schedule is marked invalid if a chosen shift length results in buffer overflow of the downstream (i.e. following) buffer. This situation can also be handled with idle time if a shop would otherwise produce more units than the downstream buffer can hold. For example: if the downstream buffer of a shop can only hold five more units before being full, the shop can still be scheduled to run for more than five hours without causing the schedule to be marked invalid. The shop will then be idle after it has produced five units. As a result, the total output volume can exhibit a complex relationship with the number of hours worked by the shops.

In the case where there are multiple shops and intermediate buffers, the calculation of the total output volume and buffer occupation becomes yet more complicated. For instance, on a given day the occupancy of the final buffer in a production line depends on the occupancy of all upstream buffers on the previous day, and on the hours assigned to the upstream shops on the current day. Not only does this imply a larger computational runtime, but it also further complicates the relationship between the total output volume and the number of hours worked, and hence the total operating cost. 

A second example is that of shared buffers, where, for instance, the output of two body shops is passed into a single buffer. The two body shops produce different vehicle models, each of which has its own total output volume target. If a single paint shop is responsible for work on both of these vehicle models, then we must specify the quantities of the different models to be passed from the buffer to the paint shop. As an example, one could choose a ratio based on the respective output volume targets. However, such a simple rule could lead to inefficiencies, since it does not take account of the actual quantities of each model in the buffer at a given time, and as a result idle time may be introduced. One could introduce more complex rules for extracting vehicles from the buffer, at the expense of a larger computational runtime.

Complications such as the two mentioned here, which are a result of more elaborate production lines, may affect the scope for restricting the search space in the spirit of the example given in \cref{app:sol_space_structure} above. 


\section{Gate requirements} \label{app:gatereqs}
In this appendix, we show the gate requirements for each high-level operator used in QISS at different levels of gate decomposition. We use this to find the total number of elementary gates required by QISS for scheduling a single day without a cost constraint, this corresponds to the circuit at the end of \cref{subsec:completedcircuit}.

\subsection{High-level operators}
The number of elementary gates for each of the high-level operators used in QISS is shown in \cref{tab:gatespergateforSSSM}. Gates are grouped by type: all types of 1-qubit gates, multi-controlled Pauli Z or X gates, double-controlled phase gates, (single) controlled phase gates, double-controlled X (Toffoli) gates, and controlled X gates. The components correspond to the components outlined before. 

Some of the components require 3-qubit gates, which can be decomposed into 1- and 2-qubit gates. The controlled U(X), the S(X) are both implemented with CCPhase gates, which can be decomposed into 3 CPhase and 2 CNOT gates each. The $\mathit{MAX}(0,\tilde{B})$ component is implemented with CCNOT (also called Toffoli) gates, which can be decomposed into 9 1-qubit gates and 6 CNOTs each. Gate totals for these components with these decompositions are also shown in the table. The diffuser component requires a multi-controlled Z gate, which can also be decomposed into 1- and 2-qubit gates. The number of such gates depends on the specific decomposition method and on the number of control qubits. For this reason, the gate has been left as-is, and no decomposed gate count is provided. 

\begin{table}
\centering
\caption{Number of elementary gates required for each of the high-level operators used in QISS. The types of elementary gates here are all types of 1-qubit gates, multi-controlled Pauli Z or X gates (MCZ/MCX), double-controlled phase gates (CCPhase), (single) controlled phase gates (CPhase), double-controlled NOT (Toffoli) gates (CCNOT), and controlled NOT gates (CNOT).}
\label{tab:gatespergateforSSSM}
\setlength{\tabcolsep}{2pt}
\begin{tabular}{lcccccc} \toprule
\begin{tabular}[c]{@{}l@{}}Circuit\\component\end{tabular}    & \begin{tabular}[c]{@{}l@{}}1-qubit\\gates \end{tabular} & \begin{tabular}[c]{@{}l@{}}MCZ/\\MCX\end{tabular}& CCPhase & CPhase & CCNOT & CNOT \\ \midrule
Output init.     & 1                  &           &         &        &       &      \\
Superposition for shops   & 2                  &           &         &        &       &      \\
Superposition for buffer & 5                  &           &         &        &       &      \\
CU(X)           &                    &           & 5       &        &       &      \\
Decomposition of CU(X)    &                    &           &         & 15     &       & 10   \\
U(X)            & 5                  &           &         &        &       &      \\
S(X)            & 4                  &           & 20      &        &       &      \\
Decomposition of S(X)     & 4                  &           &         & 60     &       & 40   \\
QFT             & 5                  &           &         & 10     &       &      \\
IQFT            & 5                  &           &         & 10     &       &      \\
$MAX(0,\tilde{B})$        &                    &           &         &        & 4     & 6    \\
Decomp. of $MAX(0,\tilde{B})$ & 36                 &           &         &        &       & 30   \\
Diffuser        & 16                 & 1         &         &        &       &  \\   \bottomrule
\end{tabular}
\end{table}

\subsection{Gate requirements for QISS}
We used the totals in \cref{tab:gatespergateforSSSM} to find the gate requirements for QISS for scheduling one day and without the cost constraint, as it is at the end of \cref{subsec:completedcircuit}. This was used as a verification step when developing the algorithm. We have included it as a companion to the qubit requirements of QISS in \cref{subsec:qubitregs}. The total gate requirement can be found in \cref{tab:totalnumberofgatesSSSM}. 

The oracle for QISS is split up into its components:
\begin{itemize}
    \item The initialization requires a single qubit (Hadamard or general initialization gate) for the shop qubits, the buffer qubits and the output qubit. 
    \item Some addition operations could be merged, so the final circuit for the oracle for the simplified model has 4 adders: $B_{init}$, $-(B_{max}+1)$, the combined $(B_{max}+1)-B_{init} + (V_{low}-1)$ and $2\Delta + 1$, each requiring 5 1-qubit phase gates.
    \item The oracle has 3 S(X) operations: $S1$, $-S2$ and $-S1$, each implemented with 4 1-qubit X-gates and 20 CCPhase gates.
    \item The $MAX(0,\tilde{B})$ operation is implemented with 4 CCNOTs and 6 CNOTs, with a leading Inverse QFT (IQFT) and followed by a QFT operation, each implemented with 5 1-qubit Hadamard gates and 10 CPhase gates. 
    \item The $c_1$ and $c_2$ constraints (maximum buffer content and minimum output volume), are each implemented with a leading IQFT, a single CNOT and followed by a QFT. In total, this is 10 1-qubit Hadamard gates, 20 CPhase and 1 CNOT gate for both conditions.
    \item The final $c_3$ condition requires only a leading IQFT, because we do not need to do any further calculations on the buffer qubits, and it therefore does not need to be "reset" as part of the oracle. In addition to the IQFT, it is implemented with a 1-qubit X gate and a CNOT, for a total of 6 1-qubit gates, 10 CPhase gates and a CNOT.
    \item In total, the oracle is thus implemented with 78 1-qubit gates, 60 CCPhase gates, 70 CPhase gates, 4 CCNOTs and 9 CNOTs.
\end{itemize}  

The complete implementation of Grover's algorithm for the simplified model consists of the following components:
\begin{itemize}
    \item The oracle, as above, which includes initialization.
    \item A multi-controlled X gate to set the output qubit based on the result of the condition qubits.
    \item The inverse of the oracle to reset the circuit, which requires the same number of gates as the oracle.
    \item Grover's diffusion operator, which consists of 2 X-gates and 2 Hadamard gates per shop qubit, and one multi-controlled Z gate over all shop qubits. This totals $2*4+2*4=16$ 1-qubit gates and 1 MCZ gate.
    \item And finally, a measurement on each of the shop qubits. 
    \item In total, the implementation of Grover's algorithm for the simplified model (for a single iteration) requires 176 1-qubit gates, an MCX and an MCX gate, 120 CCPhase gates, 140 CPhase gates, 8 CCNOT gates and 18 CNOT gates.
\end{itemize}

\begin{table}[htbp]
\centering
\caption{Total number of elementary gates required for Grover's algorithm for QISS for scheduling one day without a cost constraint, with the number of gates per component as in \cref{tab:gatespergateforSSSM}.} \label{tab:totalnumberofgatesSSSM}
\setlength{\tabcolsep}{2pt}
\begin{tabular}{@{}p{0.3cm}lcccccc@{}} \toprule
\multicolumn{2}{l}{\begin{tabular}[c]{@{}l@{}}Circuit\\component\end{tabular}}    & \begin{tabular}[c]{@{}l@{}}1-qubit\\gates \end{tabular} & \begin{tabular}[c]{@{}l@{}}MCZ/\\MCX\end{tabular} & CCPhase & CPhase & CCNOT & CNOT \\ \midrule
\multicolumn{2}{l}{Oracle}     & 78 & 0  & 60 & 70 & 4 & 9 \\ \midrule
 & Initialization       & 10          &            &             &             &            &            \\
& B$_{init}$                & 5           &            &             &             &            &            \\
& S1                   & 4           &            & 20          &             &            &            \\
& -S2                  & 4           &            & 20          &             &            &            \\
& IQFT                 & 5           &            &             & 10          &            &            \\
& $MAX(0,\tilde{B})$             &             &            &             &             & 4          & 6          \\
& QFT                  & 5           &            &             & 10          &            &            \\
& -(B$_{max}$+1)            & 5           &            &             &             &            &            \\
& $c_1$                   & 10          &            &             & 20          &            & 1          \\
& B$_{max}$-B$_{init}$+$V^*$..    & 5           &            &             &             &            &            \\
& -S1                  & 4           &            & 20          &             &            &            \\
& $c_2$                   & 10          &            &             & 20          &            & 1          \\
& 2$\Delta$-1                 & 5           &            &             &             &            &            \\
& $c_3$                   & 6           &            &             & 10          &            & 1          \\ \midrule
\multicolumn{2}{l}{Set output}  &    & 1 &    &    &   &   \\ 
\multicolumn{2}{l}{Oracle$^\dagger$} & 78 &   & 60 & 70 & 4 & 9 \\ 
\multicolumn{2}{l}{Diffuser}    & 16 & 1 &    &    &   &   \\ 
\multicolumn{2}{l}{Measurement} & 4           &            &             &             &            & \\ \midrule                           
\multicolumn{2}{l}{{Total:}} & {176} & {2} & {120} & {140} & {8} & {18} \\ \bottomrule
\end{tabular}
\end{table}

\EOD
\end{document}